\documentclass[11pt]{article}
\usepackage{amsmath}
\usepackage{natbib}
\usepackage{latexsym, epsfig, amssymb, amsmath, amsthm, graphicx, mathrsfs, booktabs}
\usepackage{rotating, tabularx, setspace,paralist}
\usepackage{bm}
\usepackage{multirow,multicol}
\usepackage{color}
\usepackage{graphicx}
\usepackage[OT1]{fontenc}
\usepackage{hypernat}
\usepackage{lscape}
\usepackage{diagbox}
\usepackage{anysize}
\usepackage{algorithm}
\usepackage{algorithmic}

\renewcommand{\baselinestretch}{1.5}
\renewcommand{\theequation}{\arabic{section}.\arabic{equation}}
\marginsize{1.in}{1in}{0.55in}{1.55in} %

\makeatletter
\def\singlespace{\def\baselinestretch{1}\@normalsize}

\newcommand{\be}{\begin{eqnarray}}
\newcommand{\ee}{\end{eqnarray}}
\newcommand{\bs}{\begin{eqnarray*}}
\newcommand{\es}{\end{eqnarray*}}

\newtheorem{assumption}{Assumption}[section]

\newtheorem{theorem}{Theorem}[section]

\newtheorem{remark}{Remark}[section]

\@addtoreset{equation}{section}

\makeatother
\allowdisplaybreaks

\begin{document}

\title{Determining The Number of Factors in Two-Way Factor Model of High-Dimensional Matrix-Variate
Time Series: A White-Noise based Method for Serial Correlation Models%
\date{}
\author{Qiang Xia$^{a}$\thanks{The author is grateful to Prof. Lixing Zhu, Prof. Xianyang Zhang and Dr. Zhaoxing Gao for many helpful comments and suggestions on an earlier version. First version:2024-1-18; This version: 2024-12-1.} 
\medskip\\
{\it \small $^a$College of Mathematics and Informatics, South China Agricultural University}\\
}
}
\maketitle
	\begin{abstract}
	In this paper, we study a new two-way factor model for high-dimensional matrix-variate time series. To estimate the number of factors in this two-way factor model, we decompose the series into two parts: one being a non-weakly correlated series and the other being a weakly correlated noise. By comparing the difference between two series, we can construct white-noise based signal statistics to determine the number of factors in row loading matrix (column loading matrix).
       Furthermore, to mitigate the negative impact on the accuracy of the estimation, which is caused by the interaction between the row loading matrix and the column loading matrix, we propose a transformation so that the transformed model only contains the row loading matrix (column loading matrix). We define sequences of ratios of two test statistics as signal statistics to determine the number of factors and derive the consistence of the estimation.
       We implement the numerical studies to examine the performance of the new methods.
	\end{abstract}
	
	
	
	\section{Introduction}\label{sec:Intro}
	
The analysis of high-dimensional time series data is becoming increasingly critical in the era of big data. 
For example, time-based data from census prediction and economic studies involve the analysis of high-dimensional time series using large-panel data. Monitoring environmental indices across various locations has further contributed to the growing high-dimensionality of environmental time series data. To address the complex structure of high-dimensional time series, factor models are commonly employed to provide a low-dimensional, parsimonious representation. For further reference, see, for vector-valued data, \cite{Pena1987}, \cite{Forni2000}, \cite{BaiNg2002}, \cite{Stock_Watson2002}, \cite{Bai2003}, \cite{Forni2005}, \cite{Hallin2007}, \cite{Pan_Yao2008}, \cite{Tsai_Tsay2010}, \cite{Lam2011}, \cite{Lam_Yao2012}, \cite{Ahn_Horenstein2013}, \cite{Fan2013}, \cite{Chang2015}, and \cite{Fan2022}.

In some scientific research fields such as economics, meteorology, and ecology,  high-dimensional matrix-variate time series have gained attention.  Despite their growing prevalence, the development of statistical methods for analyzing large-scale matrix-form data remains less explored. To achieve the model parsimony, \cite{Wang2019} introduced matrix-variate factor models that preserve the matrix structure of the data, enabling dimension reduction in both row and column directions. \cite{Chen2020} further generalized these models by incorporating domain or prior knowledge through linear constraints. In contrast to the white noise idiosyncratic error matrices studied in these two papers, \cite{Yu2021}, \cite{Chen2021}, and \cite{He2023} studied matrix-valued factor models with correlated idiosyncratic errors, where the correlations may be weak between rows, columns, or entries.
	
Under these two different model structures,  the researchers investigated the estimation of factor loading matrices and used the ratios of their eigenvalues to identify the unknown number of factors.
Specifically, \cite{Wang2019} and \cite{Chen2020} assumed that the noise matrix follows white noise, using its autocovariance structure that disregards contemporaneous correlations for estimation procedures, which allows strong contemporaneous correlations among the elements of white noise. In contrast, \cite{Yu2021} and \cite{Chen2021} employed concurrent covariance for their estimation procedures, allowing weak temporal correlations while assuming weak correlations among the elements of the noise matrix. However, when concurrent covariance is used to estimate the loading matrix as well as the number of factors in the factor model, weak correlations between the elements of the noise matrix need to be allowed. Otherwise, autocovariance is only a selection to infer the two-way factor model for the non-weakly correlated scenario between the elements of the noise matrix.
 Meanwhile, continuing the line of \cite{Wang2019}, the idiosyncratic errors can be weakly correlated.
\cite{Wang2019} suggested the eigenvalue-ratio (ER) technique to determine the number of row and column factors. However, ER-based estimation is not consistent in theory, and our numerical studies also showed that the estimation is often less accurate, particularly when the matrix-variate time series exhibits weak serial correlations or contains weak factors. As we know,  estimation inaccuracy can significantly affect downstream analysis.
	
Therefore, we propose a novel method  to estimate the number of factors in row and column with more accuracy. Taking the factor model of \cite{Wang2019} as an example, we do not directly work on the original time series, while relying on the decomposition of matrix-variate time series: a dynamic component driven by a lower-dimensional matrix-variate time series (the factor process) and a static component consisting of matrix-variate white noise. Since the white noise series exhibits no serial correlations, the decomposition is unique, meaning that for a given finite sample size,  the dimension of the factor process  and the factor loading spaces can be identified. That is, we can transfer the problem for models with serial correlation to the problem for models without serial correlation and the estimation could be easier. This decomposition facilitates both model identification and statistical inference.	
As the estimation procedures can be the same, we consider the row-loading matrix in the following. The detailed estimation for the column loading matrix is omitted in the theoretical investigation. Let $\bm{R}$ be the semi-orthogonal row-loading matrix, and $\bm{R}^\dag$ its orthogonal complement. The matrix-variate time series projected along the columns of $\bm{R}$ forms a non-white noise sequence, while its projection along the columns of $\bm{R}^\dag$ results in a white noise process. See (\ref{Model2.1}) and (\ref{Model2.2}) in the Appendix for details.  These observations are of importance because we can determine the number of factors through efficiently examining whether the projected series is white noise series. When idiosyncratic errors are weakly correlated, this novel
approach is still effective between non-weakly correlated series and weakly correlated noise.
	
	
In this paper, for the new factor model with weakly correlated idiosyncratic errors, we use the maximum absolute autocovariances and the sum of squared autocovariance matrices of the component series (after projecting onto the estimated factor loading spaces)  to construct respectively two sequences of ratios as  signal functions. Their empirical versions are defined as the signal statistics. The estimators are the maximizers that are greater than threosholds for the signal statistics. 
We also note that, based on empirical findings,  the strengths of the row and column factors interact with each other through the structure of the matrix factor model. This deteriorates  the performance of the respective estimators of the number of row and column factors. To improve accuracy of the estimation, we make, when we focus on the row-loading matrix, a transformation for the model under study  so that the transformed model only has a row  loading matrix. Afterward, we identify the number of factors using the ratio estimation, which is proved to be consistent. The same principle applies to the transformed model only having column load matrix.  Our numerical studies indicate the advantages of the transformation method
when the matrix-variate time series exhibits weak serial correlations or contains weak factors.
	
	The remainder of the paper is organized as follows. Section 2 introduces matrix-variate factor models and the new ratio-based estimators for the number of factors in rows and columns. Section 3 examines the theoretical properties of the proposed estimators. Section 4 presents simulation results and a real-data application. Section 5 concludes the research in this paper. All technical details are provided in the Appendix.
	
	Throughout the paper, let $\bm{I}_p$ denote the $p \times p$ identity matrix, and $A^\top$ denote the transpose of a matrix $A = (a_{ij})$. Define $|A|_\infty = \max_{i,j} |a_{ij}|$. Let $\|A\|_F$, $\|A\|_2$, and $\|A\|_{\min}$ denote the Frobenius norm, the maximum eigenvalue, and the minimum eigenvalue of $AA^\top$ or $A^\top A$, respectively. Let $a_n$ and $b_n$ be two sequences of numbers, and write $a_n \asymp b_n$ if $a_n = O(b_n)$ and $b_n = O(a_n)$. Suppose $A$ is a $km \times km$ square matrix with $m$ row and column partitions. Let $A_{ii}$ for $1 \leq i \leq m$ denote the $k \times k$ block matrices along the diagonal of $A$. We introduce the operation $\textmd{trs}_k(A) = \sum_{i=1}^m A_{ii}$. Finally, let $\textmd{vec}(\cdot)$ denote the vectorization operator.

	\section{Methodology}\label{sec:Model}
	\setcounter{equation}{0}
	\renewcommand{\theequation}{2.\arabic{equation}}
	\subsection{Two-way Factor Models}\label{sec:model}
	Let $\{\bm Y_t: t=1, \ldots, n\}$ represent $n$ observations from a matrix-variate time series with $\mathbb{E}\bm Y_t =\bm 0$, where $\bm Y_t$
	is a $p \times q$ matrix defined as
	\[
	\bm Y_t = \big(Y_{\cdot 1, t},\ldots, Y_{\cdot q, t}\big)
	= \begin{pmatrix}
		Y_{1\cdot, t} \\
		\vdots \\
		Y_{p\cdot, t}
	\end{pmatrix}
	= \begin{pmatrix}
		y_{11,t} & \cdots & y_{1q,t} \\
		\vdots & \ddots & \vdots \\
		y_{p1,t} & \cdots & y_{pq,t}
	\end{pmatrix}.
	\]
	\citet{Wang2019} proposed a factor model for matrix-valued time series $\bm Y_t$, given by
	\begin{equation}\label{Model1}
		\bm Y_t = \bm R \bm F_t \bm C^\top + \bm E_t, \quad t = 1, \ldots, n,
	\end{equation}
	where $\bm F_t = (F_{\cdot1,t},\ldots,F_{\cdot c,t}) = (F_{1\cdot, t},\ldots,F_{r\cdot, t})^\top$ is an unobserved $r \times c$ matrix of common fundamental factors; $\bm R$ is a $p \times r$ row loading matrix; $\bm C$ is a $q \times c$ column loading matrix; and $\bm E_t = (E_{\cdot1,t},\ldots,E_{\cdot q,t}) = (E_{1\cdot, t},\ldots,E_{p\cdot, t})^\top$ is a $p \times q$ matrix of random errors.
	In Model~\eqref{Model1}, the common fundamental factors $\bm F_t$ capture the dynamics and co-movements of $\bm Y_t$, while $\bm R$ and $\bm C$ reflect the significance of the factors and their interactions. Model~\eqref{Model1} can also be rewritten as a classical factor model:
	\begin{equation}\label{Model01}
		\text{vec}(\bm Y_t) = (\bm C \otimes \bm R) \text{vec}(\bm F_t) + \text{vec}(\bm E_t), \quad t = 1, \ldots, n,
	\end{equation}
	where $\bm C \otimes \bm R$ is the $pq \times rc$ loading matrix, with $rc$ being the total number of factors; $\text{vec}(\bm F_t)$ represents the vectorized latent factors; and $\text{vec}(\bm E_t)$ denotes the vectorized error term.
	
	For the vector-valued factor model~\eqref{Model01}, the factors $\text{vec}(\bm F_t)$ and the loading matrix $\bm C \otimes \bm R$ are identifiable only up to an invertible linear transformation. A similar identification issue arises in the matrix-valued factor model~\eqref{Model1}. Specifically, suppose $H_1$ and $H_2$ are two invertible matrices of sizes $r \times r$ and $c \times c$, respectively. Then the triplets $(\bm R, \bm F_t, \bm C)$ and $(\bm R H_1, H_1^{-1} \bm F_t H_2^{-1}, \bm C H_2^\top)$ are equivalent, implying that the factors and loadings are identifiable up to invertible transformations $H_1$ and $H_2$. However, once $\bm R$ and $\bm C$ are specified, the factor process $\bm F_t$ is uniquely determined. This non-uniqueness in $\bm R$ and $\bm C$ can be advantageous, as it allows for the selection of $H_1$ and $H_2$ to simplify estimation. For further details on identification issues in factor models, see \citet{Bai_Li2012}.
	
	\subsection{Regularity Conditions}\label{sec:R}
	To facilitate inference for the proposed model, we adopt the following regularity conditions, referring to \citet{Wang2019}.
		For $h \geq 0$, define $
		\bm\Sigma_{f}(h) = \text{Cov}\big(\text{vec}(\bm F_t), \text{vec}(\bm F_{t+h})\big),
		\bm\Sigma_{ef}(h) = \text{Cov}\big(\text{vec}(\bm E_t), \text{vec}(\bm F_{t+h})\big),
		\bm\Sigma_{fe}(h) = \text{Cov}\big(\text{vec}(\bm F_t), \text{vec}(\bm E_{t+h})\big), \bm\Sigma_{e}(h) = \text{Cov}\big(\text{vec}(\bm E_t),\text{vec}(\bm E_{t+h})\big),$ $\bm\Sigma_{e} = \text{Cov}\big(\text{vec}(\bm E_t)\big).$

	\begin{assumption}\label{assum-1}{\rm
			\begin{itemize}
				\item[\rm (C1)] There exist constants $\delta, \omega \in [0, 1]$ such that $\|\bm R\|_2^2 \asymp p^{1-\delta} \asymp \|\bm R\|_{\mathrm{min}}^2$ and $\|\bm C\|_2^2 \asymp q^{1-\omega} \asymp \|\bm C\|_{\mathrm{min}}^2$.
				\item[\rm (C2)] The vector-valued process $\big(\text{vec}^\top(\bm F_t),\text{vec}^\top(\bm E_t)\big)^\top,$ is $\alpha$-mixing, with mixing coefficients $\alpha(h)$ satisfying $\sum_{h=1}^\infty \alpha(h)^{1-2/\gamma} < \infty$ for some $\gamma > 2$, where
				\[
				\alpha(h) = \sup_{i} \sup_{A \in \mathcal{F}^i_{-\infty}, B \in \mathcal{F}^\infty_{i+h}} |P(A \cap B) - P(A)P(B)|,
				\]
				and $\mathcal{F}^j_i$ denotes the $\sigma$-field generated by $\{\big(\text{vec}^\top(\bm F_t),\text{vec}^\top(\bm E_t))^\top: i \leq t \leq j\}$.
				\item[\rm (C3)] Let $F_{t,ij}$ be the $(i,j)$-th entry of $\bm F_t$. For all $i = 1, \ldots, r$, $j = 1, \ldots, c$, and $t = 1, \ldots, n$, assume $\mathbb{E}(|F_{t,ij}|^{2\gamma}) \leq \mathcal{C}$ for some constant $\mathcal{C} > 0$. Moreover, there exists $h \in [0, h_0]$ such that $\text{rank}(\bm\Sigma_f(h)) = l = \max(r, c)$, and $\bm\Sigma_f(h) \asymp \mathbf{O}(1) \asymp \sigma_l(\bm\Sigma_f(h))$ as $p, q \to \infty$ with $r, c$ fixed.
                \item[\rm (C4)] 
                $|\bm\Sigma_{e}(h)|_\infty=\textbf{O}(n^{-1})$
                for $h\geq 1$.
				\item[\rm (C5)] The elements of $\bm\Sigma_{e}$ are bounded as $n, p, q \to \infty$. Furthermore, for $h \geq 0$, $\|\bm\Sigma_{ef}(h)\|_2 = \textbf{o}(p^{\frac{1-\delta}{2}} q^{\frac{1-\omega}{2}})$ and $\bm\Sigma_{fe}(h) = 0$.
			\end{itemize}
	}\end{assumption}
	
	Conditions (C1)-(C3) are similar to  those in \citet{Wang2019}. When $\delta = \omega = 0$, the row and column factors are referred to as strong factors. Conversely, when $\delta, \omega \neq 0$, the factors are classified as weak factors. \cite{Wang2019} assumed that the random errors are white noise, which is relaxed by Condition (C4). Condition (C4) implies that the random errors can be weak correlation, which is similar to \cite{Chen2021}. \cite{Wang2019} assumed that the random errors are not correlated with the factors.
    Condition (C5) is less restrictive, requiring only that the future white noise components are uncorrelated with the factors up to the present.

	\section{The Two-Step Procedure for Identifying The Numbers of Factors}\label{sec:Model2}
	
	\setcounter{equation}{0}
	\renewcommand{\theequation}{3.\arabic{equation}}
	Based on the structure of Model~(\ref{Model1}), the strength of the factors in the rows and columns interact with each other.  We consider a two-step procedure as follows.
	
	\subsection{Transformed Factor Models}\label{sec:model2}
	Following the approach in \cite{Yu2021}, if $\bm C$ is known, the matrix-valued time series $\bm Y_t$ can be projected to lower-dimensional spaces by setting
	\begin{eqnarray}\label{Model01}
		\bm Y_t \bm C / q^{1-\omega} = \bm R \bm F_t + \bm E_t \bm C / q^{1-\omega} := \bm R \bm F_t + \bm \xi_t.
	\end{eqnarray}
	It is easy to see that Model~(\ref{Model01}) reduces to a new one only including a row loading matrix. When the number of factors is unknown so that $\bm C$ is unknown, we choose an appropriate matrix $\widetilde{\bm C} = q^{\frac{1-\omega}{2}}(\tilde{Q}_1, \tilde{Q}_2, \ldots, \tilde{Q}_m) := (C_{\cdot 1}, \ldots, C_{\cdot m})$ with a large positive integer $m$ such that
	\[
	\bm C^\top \widetilde{\bm C} = q^{\frac{1-\omega}{2}} (\bm I_c, \bm 0),
	\]
	where $\bm I_\cdot$ denotes an identity matrix. The matrix-valued time series $\bm Y_t$ is then projected to a lower-dimensional space by setting
	\begin{eqnarray}\label{Model3.1}
		\bm X_t = \bm Y_t \widetilde{\bm C} / q^{1-\omega} = \bm R \bm F_t \bm C^\top \widetilde{\bm C} / q^{1-\omega} + \bm E_t \widetilde{\bm C} / q^{1-\omega} := \bm R \widetilde{\bm F}_t + \widetilde{\bm \xi}_t.
	\end{eqnarray}
	
	For $i, j = 1, 2, \ldots, c$, define the following covariance matrices:
	\begin{eqnarray*}
		\bm \Sigma_{x,ij}(h) &=& \frac{1}{n-h} \sum_{t=1}^{n-h} \text{Cov}(\bm Y_t C_{\cdot i}, \bm Y_{t+h} C_{\cdot j}), \\
		\bm \Sigma_{f,ij}(h) &=& \frac{1}{n-h} \sum_{t=1}^{n-h} \text{Cov}(f_{t, \cdot i}, f_{t+h, \cdot j}), \\
		\bm \Sigma_{\xi f,ij}(h) &=& \frac{1}{n-h} \sum_{t=1}^{n-h} \text{Cov}(\bm E_t C_{\cdot i}, F_{t+h, \cdot j}).
	\end{eqnarray*}
	Since $C_{\cdot i}^\top C_{\cdot j} = 0$ for $i \neq j$, and slightly differing from \cite{Wang2019}, we further define:
	\begin{eqnarray*}
		\bm \Omega_x(h) &=& \frac{1}{q^{2-2\omega}} \sum_{i=1}^{c} \bm \Sigma_{x,ii}(h) = \frac{1}{(n-h) q^{2-2\omega}} \sum_{t=1}^{n-h} \mathbb{E}(\bm Y_t \widetilde{\bm C} \widetilde{\bm C}^\top \bm Y_{t+h}^\top), \\
		\bm \Omega_f(h) &=& \sum_{i=1}^{c} \bm \Sigma_{f,ii}(h) = \frac{1}{n-h} \sum_{t=1}^{n-h} \mathbb{E}(\bm F_t \bm F_{t+h}^\top), \\
		\bm \Omega_{\xi f}(h) &=& \frac{1}{q^{1-\omega}} \sum_{i=1}^{c} \bm \Sigma_{\xi f,ii}(h) = \frac{1}{(n-h) q^{1-\omega}} \sum_{t=1}^{n-h} \mathbb{E}(\bm E_t \bm C \bm F_{t+h}^\top).
	\end{eqnarray*}
	By Assumption~\ref{assum-1}-(C5),
	\[
	\mathbb{E}(\bm R \bm F_t \bm C^\top \widetilde{\bm C} \widetilde{\bm C}^\top \bm E_{t+h}^\top) = \mathbb{E}(\bm R \bm F_t \bm C^\top \bm E_{t+h}^\top) = \text{tr}_p \left( \mathbb{E} \left( \text{vec}(\bm R \bm F_t) \text{vec}^\top(\bm E_{t+h} \bm C) \right) \right) = 0,
	\]
	and
	\[
	\mathbb{E}(\bm Y_t \widetilde{\bm C} \widetilde{\bm C}^\top \bm Y_{t+h}^\top) = q^{2-2\omega} \bm R \mathbb{E}(\bm F_t \bm F_{t+h}^\top) \bm R^\top + q^{1-\omega} \mathbb{E}(\bm E_t \bm C \bm F_{t+h}^\top) \bm R^\top.
	\]
	For a pre-determined integer $h_0 \geq 1$, we consider the main non-zero lags and define:
	\begin{eqnarray}\label{Model3.2}
		\widetilde{\bm M}_1 &=& \sum_{h=1}^{h_0} \bm \Omega_x^\top(h) \bm \Omega_x(h) = \bm R \left( \sum_{h=1}^{h_0} \left( \bm R \bm \Omega_f(h) + \bm \Omega_{\xi f}(h) \right)^\top \left( \bm R \bm \Omega_f(h) + \bm \Omega_{\xi f}(h) \right) \right) \bm R^\top.
	\end{eqnarray}
	By Assumption~\ref{assum-3}, the matrix $\widetilde{\bm M}_1$ has rank $r$. From Equation~(\ref{Model3.2}), we see that each column of $\widetilde{\bm M}_1$ is a linear combination of the columns of $\bm R$, and thus the eigenspace of $\widetilde{\bm M}_1$ is the same as the eigenspace of $\bm R$, i.e., $\mathcal{M}(\widetilde{\bm M}_1) = \mathcal{M}(\bm R)$. Therefore, $\mathcal{M}(\bm R)$ can be estimated by the space spanned by the eigenvectors of the sample version of $\widetilde{\bm M}_1$. Let $O_j$ be the unit eigenvector corresponding to the $j$-th largest eigenvalue of $\widetilde{\bm M}_1$. Define $\bm O = (O_1, O_2, \ldots, O_r)$, and then obtain $\bm R = p^{\frac{1-\delta}{2}} \bm O$.
	
	We construct the sample versions of these quantities to estimate the factor loading matrix as follows.
	Let
	\begin{eqnarray}\label{Model3.3}
		\widehat{\widetilde{\bm M}}_1= \sum_{h=1}^{h_0}\widehat{\bm\Omega}^\top_{x}(h)\widehat{\bm\Omega}_{x}(h),
	\end{eqnarray}
	where $\widehat{\bm\Omega}_{x}(h)=\frac{1}{(n-h)q^{2-2\omega}}\sum_{t=1}^{n-h}\bm Y_t\widehat{\widetilde{\bm C}}\widehat{\widetilde{\bm C}}^\top\bm Y^\top_{t+h}$.
	Then, $\mathcal{M}(\bm R)$ can be estimated by $\mathcal{M}(\widehat{\bm O})$, where
	$\widehat{\bm O} = (\hat{O}_{1}, \ldots, \hat{O}_{r})$,
	and $\hat{O}_{1}, \ldots, \hat{O}_{r}$ are the unit eigenvectors of $\widehat{\widetilde{\bm M}}_1$ corresponding to its $r$ largest eigenvalues.
	As a result, $\widehat{\bm R}=p^{\frac{1-\delta}{2}}\widehat{\bm O}$.
	
	\begin{remark} First, $\widehat{\widetilde{\bm C}}$ can be estimated by (S.2) in Appendix.
		In fact, a small positive value $m$ can also be chosen. Then $\bm C^\top\widetilde{\bm C}=q^{\frac{1-\omega}{2}}(\bm I_m, \bm 0)^\top.$
		The result and the relative proof are similar. Thus, we do not discuss them further. Second, omitting to $q^{2-2\omega}$ in practice, we can consider that $\bm\Omega_{x}(h)=\frac{1}{n-h}\sum_{t=1}^{n-h}\mathbb{E}(\bm Y_t\widetilde{\bm C}\widetilde{\bm C}^\top\bm Y^\top_{t+h})$,
		and $\widetilde{\bm M}_1 = \sum_{h=1}^{h_0}\bm\Omega^\top_{x}(h)\bm\Omega_{x}(h)$ accordingly.
		Third, according to \cite{Wang2019}, the selection of $h_0$ is not sensitive to the estimations of the number of factors and the loading matrix. In this paper, we set $h_0=2$ in all simulation experiments and the application.
	Fourth, the autocovariance is used to construct statistics for estimating the loading matrix and the number of factors in rows and columns. If we employ the concurrent covariance, the additional condition, that is $\|\bm\Sigma_{e}\|_2=\textbf{o}(p^{1-\delta} q^{1-\omega})$, should be satisfied.
    \end{remark}
	
	\subsection{Max-type and sum-type estimators}\label{sec:est30}
	
	For Model~(\ref{Model3.1}), as discussed in Section~\ref{sec:model}, we can assume without loss of generality that $\bm O = (O_1, \dots, O_r) = \bm R / p^{\frac{1-\delta}{2}}$ is semi-orthogonal, i.e., $\bm O^\top \bm O = \bm I_r$, where $O_i \in \mathbb{R}^{p \times 1}$ for $1 \leq i \leq r$. Let $\bm O^\dag = (O_{r+1}, \ldots, O_p)$ be the orthogonal complement matrix of $\bm O$, so that $(\bm O, \bm O^\dag)^\top (\bm O, \bm O^\dag) = \bm I_p$. For a better understanding of our approach, we assume that $\bm E_t$ is white noise firstly.
    Then we have:
	\begin{eqnarray}
		O_i^\top \bm X_t &=& O_i^\top \bm R \bm F_t \widetilde{\bm C} / q^{\frac{1-\omega}{2}} + O_i^\top \bm E_t \widetilde{\bm C} / q^{\frac{1-\omega}{2}}, \quad i = 1, \ldots, r; \label{Model3.4} \\
		O_i^\top \bm X_t &=& O_i^\top \bm E_t \widetilde{\bm C} / q^{\frac{1-\omega}{2}}, \quad i = r+1, \ldots, p. \label{Model3.5}
	\end{eqnarray}
	This implies that $O_i^\top \bm X_t$ in (\ref{Model3.4}) is not a white noise sequence for $i = 1, \ldots, r$, while $O_i^\top \bm X_t$ in (\ref{Model3.5}) is a white noise sequence for $i = r+1, \ldots, p$. This observation offers the motivation to develop a new approach to estimate the number of factors $r$ by examining the uncorrelatedness of the sequences $O_i^\top \bm X_t$ for $i = 1, \ldots, p$.
	
	Let $\mathfrak{R}_i = (O_i, \ldots, O_p)$, for $i = 1, \ldots, p$. The estimator is based on statistics for checking the white noise assumption on the sequence $\{\mathfrak{R}_i^\top \bm X_t : t = 1, 2, \ldots\}$. Define the sample covariance matrix for $\bm X_t$ as
	\[
	\bm \Sigma_X(h) := \frac{1}{n-h} \sum_{t=1}^{n-h} \mathbb{E}(\bm X_t \bm X_{t+h}^\top) \in \mathbb{R}^{p \times p}.
	\]
	The auto-covariance matrix for $\{\mathfrak{R}_i^\top \bm X_t\}$ at lag $h$ is given by
	\begin{eqnarray}\label{Model3.6}
		\breve{\Gamma}_i(h) := \Big(\breve{\gamma}_{i,kl}(h)\Big) = \mathfrak{R}_i^\top \bm \Sigma_X(h) \mathfrak{R}_i.
	\end{eqnarray}
	
    Referring to \cite{Chang2017} and \cite{Li2019}, the max-type and sum-type tests can be proposed below to identify whether the sequence $\{\mathfrak{R}_i^\top \bm X_t : t = 1, 2, \ldots\}$ is white noise or not. The max-type
    test is constructed based on the maximum norm of the auto-covariance matrix:
	\begin{eqnarray}\label{Model3.7}
		\breve{T}_{i,n} := \max_{1 \leq h \leq K} n^{1/2} \left|\breve{\Gamma}_i(h)\right|_\infty = \max_{1 \leq h \leq K} n^{1/2} \left|\mathfrak{R}_i^\top \bm \Sigma_X(h) \mathfrak{R}_i\right|_\infty,
	\end{eqnarray}
	where $K$ is a user-specified integer.
	The sum-type test is constructed based on the Frobenius norm of the auto-covariance matrix:
	\begin{eqnarray}\label{Model3.8}
		\breve{G}_{i,n} := \sum_{1 \leq h \leq K} \text{Tr}\left(\breve{\Gamma}_i(h)^\top \breve{\Gamma}_i(h)\right) = \sum_{1 \leq h \leq K} \left\|\breve{\Gamma}_i(h)\right\|_F^2 = \sum_{1 \leq h \leq K} \sum_j \breve{\sigma}_j^2\left(\breve{\Gamma}_i(h)\right),
	\end{eqnarray}
	where $\text{Tr}$ denotes the trace operation for a square matrix and $\breve{\sigma}_j^2$ denotes the $j$-th largest singular value of $\breve{\Gamma}_i(h)$. Large values of $\breve{T}_{i,n}$ and $\breve{G}_{i,n}$ indicate a potential departure from the assumption of white noise in the sequence $\{\mathfrak{R}_i^\top \bm X_t\}$. To estimate the number of factors $r$, we sequentially examine the values of the statistics as $i$ grows. Intuitively, when $i = r$, the gap between $\breve{T}_{i,n}$ and $\breve{T}_{i+1,n}$ is likely to reach its maximum value, as $\{\mathfrak{R}_{r+1}^\top \bm X_t\}$ is a white noise sequence, while $\{\mathfrak{R}_r^\top \bm X_t\}$ is a non-white noise sequence. When $\bm E_t$ is weakly correlated, the same phenomenon can be observed. It is merely that $\breve{T}_{i,n}$=$\breve{G}_{i,n}$=0 under the assumption of white noise for $\bm E_t$, while $\breve{T}_{i,n}>0$ and $\breve{G}_{i,n}>0 $ under the assumption of weak correlation for $\bm E_t$.
	
    By Lemma~\ref{lem_80},
    $\breve{T}_{i,n} $ and $\breve{G}_{i,n}$ are monotonically decreasing with respect to $i$. Based on (\ref{Model3.2}), we can construct the feasible max-type and sum-type statistics as:
	\begin{align*}
		&\widehat{\breve{T}}_{i,n} = \max_{1 \leq h \leq K} n^{1/2} \left|\widehat{\breve{\Gamma}}_i(h)\right|_\infty, \\
		&\widehat{\breve{G}}_{i,n} = \sum_{1 \leq h \leq K} \text{Tr}\left(\widehat{\breve{\Gamma}}_i(h)^\top \widehat{\breve{\Gamma}}_i(h)\right),
	\end{align*}
	where $\widehat{\breve{\Gamma}}_i(h) = \widehat{\mathfrak{R}}_i^\top \widehat{\bm \Sigma}_X(h) \widehat{\mathfrak{R}}_i$. By Lemma~\ref{lem_80}, $\widehat{\breve{T}}_{i,n}$ and $\widehat{\breve{G}}_{i,n}$ also decrease monotonically with respect to $i$. The following ratio estimators for $r$ are proposed based on an enhanced elbow criterion:
	\begin{eqnarray}
		&&\widehat{r}_{\mathrm{M R}} = \arg\max_{1 \leq i \leq p/2} \frac{\widehat{\breve{T}}_{i,n}}{\widehat{\breve{T}}_{i+1,n}} =: \arg \max_{1 \leq i \leq p/2} \mathrm{M R}(i), \label{Model3.9} \\
		&&\widehat{r}_{\mathrm{SR}} = \arg\max_{1 \leq i \leq p/2} \frac{\widehat{\breve{G}}_{i,n} - \widehat{\breve{G}}_{i+1,n}}{\widehat{\breve{G}}_{i+1,n} - \widehat{\breve{G}}_{i+2,n}} =: \arg \max_{1 \leq i \leq p/2} \mathrm{SR}(i), \label{Model3.10}
	\end{eqnarray}
	where $\mathrm{M R}(i) = \widehat{\breve{T}}_{i,n} / \widehat{\breve{T}}_{i+1,n}$ and $\mathrm{SR}(i) = (\widehat{\breve{G}}_{i,n} - \widehat{\breve{G}}_{i+1,n}) / (\widehat{\breve{G}}_{i+1,n} - \widehat{\breve{G}}_{i+2,n})$.
	

	\begin{remark} Similarly to    $R_{MR}$ and $R_{SR}$ defined in (\ref{Model3.9}) and (\ref{Model3.10}), two statistics using autocorrelations and cross-correlations were respectively  employed by \cite{Chang2018} for time series principal component analysis and by \cite{Han2022} for decorrelation bilinear transformation. For $\widehat{r}_{\mathrm{M R}}$ and $\widehat{r}_{\mathrm{S R}}$,
    if the random errors are weak correlation, both $\widehat{\breve{T}}_{i,n}$ and $\widehat{\breve{G}}_{i,n}$ are greater than zero. Then their consistency can be ensured.
     However, because $\widetilde{\bm M}_1$ is not full rank, the ER method based on $\widetilde{\bm M}_1$ is not consistent.
	\end{remark}
	
	
	Similarly, we can estimate the number of column factors $c$. As the procedure is almost the same, we only introduce the notation and estimators. Select the suitable $\widetilde{\bm R} = p^{\frac{1-\delta}{2}}(Q_1, Q_2, \dots, Q_m) = (R_{\cdot 1}, \dots, R_{\cdot m})$ with a large positive integer $m$, and define:
	\begin{eqnarray*}
		&&\bm \Omega_{\tilde{x}}(h) := \frac{1}{(n-h)p^{2-2\delta}} \sum_{t=1}^{n-h} \mathbb{E}(\bm Y_t^\top \widetilde{\bm R} \widetilde{\bm R}^\top \bm Y_{t+h}), \\
		&&\bm \Omega_{\tilde{f}}(h) := \frac{1}{n-h} \sum_{t=1}^{n-h} \mathbb{E}(\bm F_t^\top \bm F_{t+h}), \\
		&&\bm \Omega_{\tilde{\xi} \tilde{f}}(h) := \frac{1}{(n-h)p^{1-\delta}} \sum_{t=1}^{n-h} \mathbb{E}(\bm E_t^\top \bm R \bm F_{t+h}).
	\end{eqnarray*} Further, define the matrix:
	\begin{eqnarray}\label{Model3.12}
		\widetilde{\bm M}_2 = \sum_{h=1}^{h_0} \bm \Omega_{\tilde{x}}^\top(h) \bm \Omega_{\tilde{x}}(h) = \bm C \left( \sum_{h=1}^{h_0} \left( \bm C \bm \Omega_{\tilde{f}}(h) + \bm \Omega_{\tilde{\xi} \tilde{f}}(h) \right)^\top \left( \bm C \bm \Omega_{\tilde{f}}(h) + \bm \Omega_{\tilde{\xi} \tilde{f}}(h) \right) \right) \bm C^\top,
	\end{eqnarray}
	and its sample version:
	\begin{eqnarray}\label{Model3.13}
		\widehat{\widetilde{\bm M}}_2 = \sum_{h=1}^{h_0} \widehat{\bm \Omega}_{\tilde{x}}^\top(h) \widehat{\bm \Omega}_{\tilde{x}}(h),
	\end{eqnarray}
	where $\widehat{\bm \Omega}_{\tilde{x}}(h) = \frac{1}{(n-h)q^{2-2\omega}} \sum_{t=1}^{n-h} \bm Y_t^\top \widehat{\widetilde{\bm R}} \widehat{\widetilde{\bm R}}^\top \bm Y_{t+h}$.
	Let $\tilde{O}_j$ and $\hat{\tilde{O}}_j$ be the unit eigenvectors corresponding to the $j$-th largest eigenvalue of $\widetilde{\bm M}_2$ and $\widehat{\widetilde{\bm M}}_2$, respectively. We define $\widetilde{\bm O} = (\tilde{O}_1, \tilde{O}_2, \dots, \tilde{O}_c)$ and $\widehat{\widetilde{\bm O}} = (\hat{\tilde{O}}_1, \hat{\tilde{O}}_2, \dots, \hat{\tilde{O}}_c)$, and then obtain $\bm C = q^{\frac{1-\omega}{2}} \widetilde{\bm O}$ and $\widehat{\bm C} = q^{\frac{1-\omega}{2}} \widehat{\widetilde{\bm O}}$.
	
	Let $\mathfrak{C}_i = (\tilde{O}_i, \dots, \tilde{O}_q)$ for $i = 1, \dots, q$, and define the covariance matrix for $\widetilde{\bm X}_t$ as:
	\[
	\bm \Sigma_{\widetilde{X}}(h) := \frac{1}{n-h} \sum_{t=1}^{n-h} \mathbb{E}(\widetilde{\bm X}_t \widetilde{\bm X}_{t+h}^\top) \in \mathbb{R}^{q \times q}.
	\]
	The auto-covariance matrix for $\{\mathfrak{C}_i^\top \widetilde{\bm X}_t\}$ at lag $h$ is:
	\begin{eqnarray}\label{Model3.14}
		\check{\Gamma}_i(h) := \mathfrak{C}_i^\top \bm \Sigma_{\widetilde{X}}(h) \mathfrak{C}_i.
	\end{eqnarray}
	The max-type test statistic is given by:
	\begin{eqnarray}\label{Model3.15}
		\check{T}_{i,n} := \max_{1 \leq h \leq K} n^{1/2} \left| \check{\Gamma}_i(h) \right|_\infty = \max_{1 \leq h \leq K} n^{1/2} \left| \mathfrak{C}_i^\top \bm \Sigma_{\widetilde{X}}(h) \mathfrak{C}_i \right|_\infty,
	\end{eqnarray}
	and the sum-type test statistic is:
	\begin{eqnarray}\label{Model3.16}
		\check{G}_{i,n} := \sum_{1 \leq h \leq K} \text{Tr} \left( \check{\Gamma}_i^\top(h) \check{\Gamma}_i(h) \right) = \sum_{1 \leq h \leq K} \left\| \check{\Gamma}_i(h) \right\|_F^2 = \sum_{1 \leq h \leq K} \sum_j \check{\sigma}_j^2 \left( \check{\Gamma}_i(h) \right),
	\end{eqnarray}
	where $\check{\sigma}_j^2$ denotes the $j$-th largest singular value of $\check{\Gamma}_i(h)$.
	
	The sample versions of the max-type and sum-type statistics are:
	\begin{align*}
		&\widehat{\check{T}}_{i,n} = \max_{1 \leq h \leq K} n^{1/2} \left| \widehat{\check{\Gamma}}_i(h) \right|_\infty, \\
		&\widehat{\check{G}}_{i,n} = \sum_{1 \leq h \leq K} \text{Tr} \left( \widehat{\check{\Gamma}}_i^\top(h) \widehat{\check{\Gamma}}_i(h) \right),
	\end{align*}
	where $\widehat{\check{\Gamma}}_i(h) = \widehat{\mathfrak{C}}_i^\top \widehat{\bm \Sigma}_{\tilde{X}}(h) \widehat{\mathfrak{C}}_i$.
	The ratio-based estimators for the number of column factors $c$ are:
	\begin{eqnarray}
		&&\widehat{c}_{\mathrm{MR}} = \arg \max_{1 \leq i \leq q/2} \frac{\widehat{\check{T}}_{i,n}}{\widehat{\check{T}}_{i+1,n}} =: \arg \max_{1 \leq i \leq q/2} \mathrm{M R}(i), \label{Model3.17} \\
		&&\widehat{c}_{\mathrm{SR}} = \arg \max_{1 \leq i \leq q/2} \frac{\widehat{\check{G}}_{i,n} - \widehat{\check{G}}_{i+1,n}}{\widehat{\check{G}}_{i+1,n} - \widehat{\check{G}}_{i+2,n}} =: \arg \max_{1 \leq i \leq q/2} \mathrm{SR}(i), \label{Model3.18}
	\end{eqnarray}
	where $\mathrm{M R}(i) = \frac{\widehat{\check{T}}_{i,n}}{\widehat{\check{T}}_{i+1,n}}$, and $\mathrm{SR}(i) = \frac{\widehat{\check{G}}_{i,n} - \widehat{\check{G}}_{i+1,n}}{\widehat{\check{G}}_{i+1,n} - \widehat{\check{G}}_{i+2,n}}$.
	
Finally, we suggest the following Algorithm~\ref{alg1} to specify the numbers of row and column factors.
	\begin{algorithm}[H]
		\caption{Two-step procedure for specifying the numbers of factors}\label{alg1}
		\textbf{Input:} Data matrices $\{\bm Y_t\}_{t \leq T}$, maximum number $m$\\
		\textbf{Output:} Numbers of row and column factors $\widehat{r}_{\mathrm{M R}}$, $\widehat{r}_{\mathrm{S R}}$, $\widehat{c}_{\mathrm{M R}}$, $\widehat{c}_{\mathrm{S R}}$
		\begin{algorithmic}[1]
			\STATE Initial estimators $\widehat{\widetilde{\bm R}}$ and $\widehat{\widetilde{\bm C}}$ from $\widehat{\bm M}_1$ and $\widehat{\bm M}_2$;
			\STATE \begin{itemize}
				\item Calculate $\widehat{\widetilde{\bm M}}_1$ using $\widehat{\widetilde{\bm C}}$, then obtain $\widehat{r}_{\mathrm{M R}}$ and $\widehat{r}_{\mathrm{S R}}$ from equations (\ref{Model3.9}) and (\ref{Model3.10});
                \item Calculate $\widehat{\widetilde{\bm M}}_2$ using $\widehat{\widetilde{\bm R}}$, then obtain $\widehat{c}_{\mathrm{M R}}$ and $\widehat{c}_{\mathrm{S R}}$ from equations (\ref{Model3.17}) and (\ref{Model3.18}).
            \end{itemize}
		\end{algorithmic}
	\end{algorithm}

	\subsection{Asymptotic Properties of The Estimators }\label{sec:theory1}
	The properties of the proposed estimators for the two-step method are investigated under the asymptotic region where $p,q$ and $n$ all tend to infinity while $r$ and $c$ are fixed. The rates of convergence for the ratios of the max-type and sum-type statistics as well as the consistency of the proposed estimators are established
	in the following theories.
	
	\begin{assumption}\label{assum-3}{\rm
			$\widetilde{\bm M}_1$ has $r$ distinct positive eigenvalues; $\widetilde{\bm M}_2$ has $c$ distinct positive eigenvalues.
		}
	\end{assumption}
	\begin{theorem}\label{theo4} Under Assumptions~\ref{assum-1} and ~\ref{assum-3} 
		and $p^{\delta}q^{\omega}n^{-1/2}=\textbf{o}(1)$, we have
		\begin{eqnarray*}
			&&\dfrac{\widehat{\breve{T}}_{r+1,n}}{\widehat{\breve{T}}_{r,n}}=\textbf{O}_p\big(p^{2\delta}q^{2\omega}n^{-1}\big);
			\quad\dfrac{\widehat{\check{T}}_{c+1,n}}{\widehat{\check{T}}_{c,n}}=\textbf{O}_p\big(p^{2\delta}q^{2\omega}n^{-1}\big).
		\end{eqnarray*}
	\end{theorem}
	\begin{theorem}\label{theo5} Under Assumptions~\ref{assum-1} and~\ref{assum-3} 
		and $p^{\delta}q^{\omega}n^{-1/2}=\textbf{o}(1)$, we have
		\begin{eqnarray*}
			&&\dfrac{\widehat{\breve{G}}_{r+1,n}-\widehat{\breve{G}}_{r+2,n}}{\widehat{\breve{G}}_{r,n}-\widehat{\breve{G}}_{r+1,n}}=\textbf{O}_p\big(p^{2\delta}q^{2\omega}n^{-1}\big);\quad
			\dfrac{\widehat{\check{G}}_{c+1,n}-\widehat{\check{G}}_{c+2,n}}{\widehat{\check{G}}_{c,n}-\widehat{\check{G}}_{c+1,n}}=\textbf{O}_p\big(p^{2\delta}q^{2\omega}n^{-1}\big).
		\end{eqnarray*}
	\end{theorem}
    	\begin{theorem}\label{theo40} Suppose Assumptions~\ref{assum-1} and~\ref{assum-3} hold 
		and $p^{\delta}q^{\omega}n^{-1/2}=\textbf{o}(1)$. Then, the $\mathrm{MR}$, $\mathrm{SR}$, $\mathrm{\widetilde{MR}}$ and $\mathrm{\widetilde{SR}}$ estimators (i.e., $\widehat{r}_{\mathrm{MR}}$ in (\ref{Model3.9}), $\widehat{r}_{\mathrm{SR}}$ in (\ref{Model3.10}), $\widehat{c}_{\mathrm{M R}}$ in (\ref{Model3.17}) and $\widehat{c}_{\mathrm{S R}}$ in (\ref{Model3.18}))
        satisfy that
		\begin{eqnarray*}
			&&P(\widehat{r}_{\mathrm{SR}}=r)\to 1; \quad P(\widehat{c}_{\mathrm{SR}}=c)\to 1;\\
			&&P(\widehat{r}_{\mathrm{MR}}=r)\to 1; \quad P(\widehat{c}_{\mathrm{MR}}=c)\to 1.
		\end{eqnarray*}
	\end{theorem}
	\begin{remark}
		First, when $\delta=\omega<0.5$ and $n\asymp pq$, the condition for weak factors is $p^{\delta}q^{\omega}n^{-1/2}=\textbf{o}(1)$; when $\delta=\omega=0$, the convergence rates for strong factors are all $n^{-1}$.
		Second, all proposed ratios are the smallest in position $r$ or $c$, which is just the number of factors in the row or column.
		From Theorem~\ref{theo5} and Theorem~\ref{theo4}, the two-step estimators can achieve the same convergence rate as the counterparts in Theorem~\ref{theo3} and Theorem~\ref{theo2} of the Appendix. 
	\end{remark}
	
	
	
	\section{Numerical Studies}
	\setcounter{equation}{0}
	\renewcommand{\theequation}{4.\arabic{equation}}
	
	\subsection{Simulation Experiments}
	
	We conduct simulation experiments to compare the proposed method with the eigenvalue ratio (ER) method suggested by \cite{Wang2019}. Specifically, we simulate $\bm Y_t$'s from model (\ref{Model1}), where the dimensions of the latent factor process $\bm F_t$ are chosen to be $r = c = 3$. The entries of $\bm F_t$ are independent and follow AR(1) processes, with errors generated from the standard normal distribution. That is:
	\[
	F_{ij,t} = a_{ij} F_{ij,t-1} + \varepsilon_{ij,t}, \quad i = 1, \ldots, r; \quad j = 1, \ldots, c; \quad t = 1, \ldots, n,
	\]
	where $a_{ij} = \omega_{ij} a$, with $\omega_{ij}$ being i.i.d. random variables such that $P(\omega_{ij} = \pm 1) = 0.5$. We consider three scenarios for $a$:
	\begin{itemize}
		\item [\rm (1)] Weak correlations: $a = 0.1$;
		\item [\rm (2)] Moderate correlations: $a = 0.5$;
		\item [\rm (3)] Strong correlations: $a = 0.9$.
	\end{itemize}
	
	The entries of $\bm R$ and $\bm C$ are independently sampled from the uniform distribution on $(-1, 1)$. Following \cite{Wang2019}, we consider the idiosyncratic errors $\bm E_t = \bm \Sigma_1^{1/2} \bm \varepsilon_t \bm \Sigma_2^{1/2}$, where $\bm \varepsilon_t$ follows a normal distribution with $\mathbb{E}(\bm \varepsilon_t) = 0$ and $\text{Cov}(\text{vec}(\bm \varepsilon_t)) = \bm I_{pq}$.
	
	We explore the following two cases for the covariance matrices $\bm \Sigma_1$ and $\bm \Sigma_2$:
	\begin{itemize}
		\item [\rm (1)] $\bm \Sigma_1 = \bm I_p$ and $\bm \Sigma_2 = \bm I_q$;
		\item [\rm (2)] $\bm \Sigma_1 = (\bm \sigma_{1,ij})$, where $\sigma_{1,ij} = 0.1$ for $i \neq j$ and $\sigma_{1,ii} = 1$, and $\bm \Sigma_2 = (\bm \sigma_{2,ij})$, where $\sigma_{2,ij} = 0.1$ for $i \neq j$ and $\sigma_{2,ii} = 1$.
	\end{itemize}
	
	We consider three combinations of $(\delta, \omega)$, namely $(0.5, 0.5)$, $(0.5, 0)$, and $(0, 0)$. For each combination of $\delta$ and $\omega$, the dimensions $(p, q)$ are chosen as $(20, 20)$, $(20, 40)$, and $(40, 40)$. The sample size $n$ is set to either 200 or 800. Since $K \geq 2$, the proposed estimators perform well, and we choose $K = 3$ for the simulations and one application in this paper. The simulation results are based on 200 Monte Carlo replications.
	
	We report the relative frequencies of occurrences where $\hat{r} = r$ or $\hat{c} = c$ (denoted by $x$), the frequency of $\hat{r} < r$ or $\hat{c} < c$ (denoted by $y$), and the frequency of $\hat{r} > r$ or $\hat{c} > c$ (denoted by $z$) among the 200 replications in Tables~\ref{tab1}-\ref{tab6}, where the results are presented in the form $x(y|z)$. Here, the estimators $ER_o$, $SR_o$, and $MR_o$ are defined based on model~(\ref{Model1}), as detailed in Appendix.
	
	Tables~\ref{tab1}-\ref{tab3} show the case where $\bm \Sigma_1$ and $\bm \Sigma_2$ are diagonal. Based on model~(\ref{Model1}), the estimated results demonstrate that the proposed approaches perform excellently. The $SR_o$ estimator is the most powerful method for determining the number of factors in both rows and columns among the three estimators, while the $ER_o$ estimator is less effective in this scenario. Moreover, all estimators based on the transformed model~(\ref{Model3.1}) perform better than their counterparts based on model~(\ref{Model1}), especially when the latent factors exhibit weak correlations. Both $SR$ and $MR$ estimators consistently yield better results than the $ER$ estimator.
	
	When the elements in $\bm \Sigma_1$ and $\bm \Sigma_2$ are small but equal, this case illustrates that the elements of $\bm E_t$ are weakly correlated. Based on model~(\ref{Model1}), Tables~\ref{tab4}-\ref{tab6} show that the $ER_o$ estimator performs poorly, and the estimators $SR_o$ and $MR_o$ do not perform satisfactorily under weak factors and small correlations. However, based on the transformed model~(\ref{Model3.1}), the proposed estimators achieve significantly better results than the $ER$ estimator. Specifically, the $SR$ estimator consistently outperforms the others, with its improved power being most notable. At the same time, the $MR$ estimator clearly outperforms the $ER$ estimator.
	
	In summary, Tables~\ref{tab1}-\ref{tab6} indicate that the performance of all estimators improves significantly as the strength or serial correlations of the latent factors increase, and none of the estimators is sensitive to the sample size. Furthermore, the proposed methods are more powerful than the $ER$ method in all cases based on the transformed models. Comparing the $MR$ estimator, the $SR$ estimator is more robust.
	
	\subsection{Real Data Analysis}
	
	We employ the matrix factor model to analyze a ten-by-ten return series comprising market equity (ME) and the Book-to-Market ratio (BE/ME). The dataset consists of 100 portfolios, constructed by intersecting ten portfolios based on ME and ten portfolios based on BE/ME. The analysis is performed on the daily returns of these portfolios, collected from January 6, 2021, to June 28, 2024, covering a total of 875 months. For further details, please refer to {\it http://mba.tuck.dartmouth.edu/pages/faculty/ken.french/data\_library.html}.

\begin{figure}
		\centering
		\includegraphics[width=13cm]{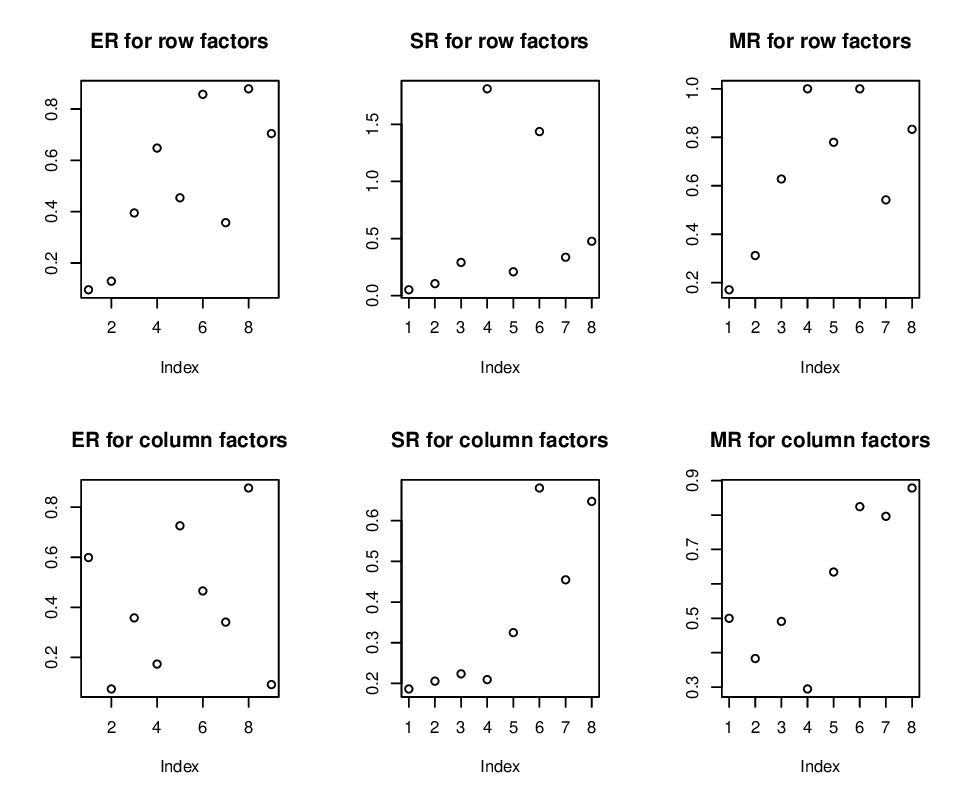}
		\caption{The number of row and column factors by $ER_o$, $SR_o$ and $MR_o$ estimators for the real data set
			\label{fig1}}
	\end{figure}

\begin{figure}
		\centering
		\includegraphics[width=13cm]{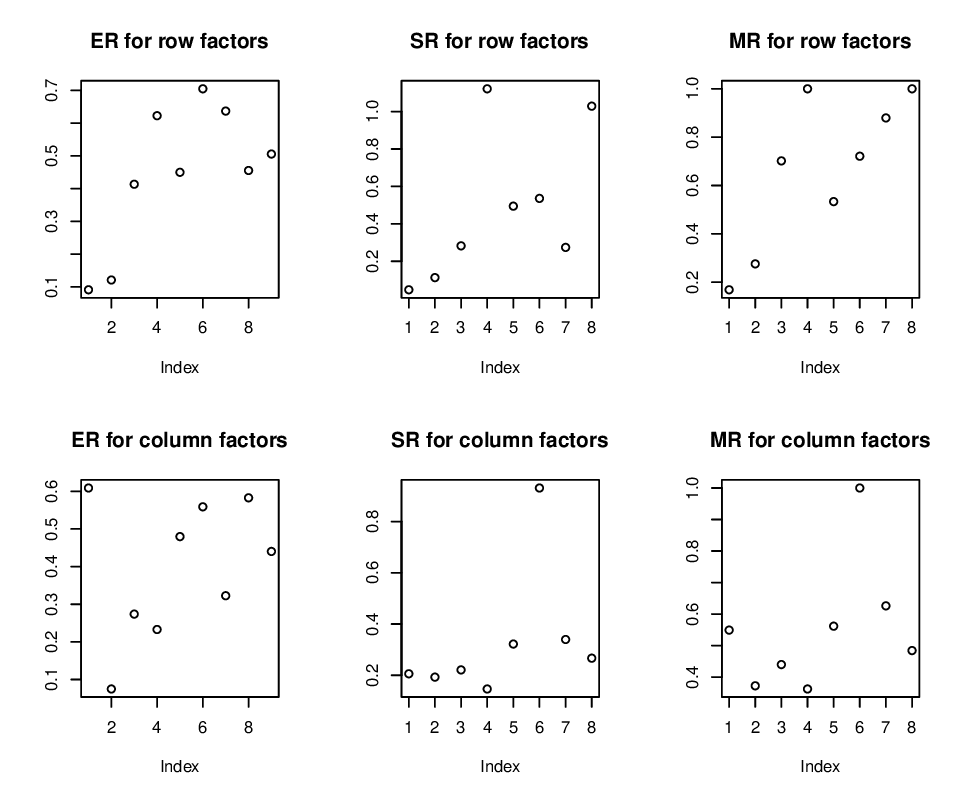}
		\caption{The number of row and column factors by ER, SR and MR estimators for the real data set
			\label{fig2}}
	\end{figure}

	Firstly, we utilize the $ER_o$, $SR_o$, and $MR_o$ methods to estimate the numbers of row and column factors in the matrix-variate factor model. As illustrated in Figure~\ref{fig1}, the $ER_o$ estimator suggests one row factor and two column factors (i.e., $r=1$, $c=2$); the $SR_o$ method indicates one row factor and one column factor (i.e., $r=1$, $c=1$); and the $MR_o$ approach implies one row factor and four column factors (i.e., $r=1$, $c=4$).
	Next, based on the transformed matrix factor model, we apply the $ER$, $SR$, and $MR$ methods to determine the numbers of row and column factors. As shown in Figure~\ref{fig2}, the $ER$ estimator yields the same results as before, i.e., $r=1$, $c=2$, while both the $MR$ and $SR$ methods indicate one row factor and four column factors (i.e., $r=1$, $c=4$).
	
	To compare models with different numbers of factors, we use $L$-fold cross-validation. Specifically, we divide the entire dataset $D$ into $L$ subsamples, $D_1, \dots, D_L$, and fit a factor model using each of the $D_{-l}$ samples to obtain $\widehat{\bm R}_{-l}$ and $\widehat{\bm C}_{-l}$, where $D_{-l}$ denotes the dataset with the $l$-th subsample removed. For each data subset $D_l$, we define the dynamic signal part of the factor model as $\bm S_{t,l} = \bm R \bm F_t \bm C^\top$. The estimated loading spaces are then used to obtain $\widehat{\bm S}_{t,l} = \widehat{\bm R}_{-l} \widehat{\bm R}^\top_{-l} \bm Y_t \widehat{\bm C}_{-l} \widehat{\bm C}^\top_{-l}$ for each $l = 1, \dots, L$.
	
	Next, we compute the out-of-sample residuals $\bm Y_t - \widehat{\bm S}_{t,l}$ and calculate the adjusted residual sum of squares (RSS) over the $L$ folds for comparison:
	\[
	RSS = \frac{1}{pq} \sum_{l=1}^L \sum_{t \in D_l} \|\bm Y_t - \widehat{\bm S}_{t,l}\|_F.
	\]
	Here, we set $L = 5$ and ensure that the time length of each dataset $D_l$ is 175. We compute $RSS$ for the matrix factor model under two different scenarios: when $r = 1$ and $c = 2$, the value $RSS$ is 6.673, while when $r = 1$ and $c = 4$, the value $RSS$ is 6.455. These results indicate that both $SR$ and $MR$ estimators perform better than $ER$ estimator, suggesting that incorporating the appropriate number of factors in the matrix factor model leads to more accurate out-of-sample predictions.

	\section{Conclusion} \label{sec:Col}
	In this paper, we proposed two new ratio-based estimators for determining the number of row and column factors in two-way high-dimensional matrix-variate time series factor models. These estimators were developed based on the element-wise maximum norm and the Frobenius norm of the sample auto-covariance matrices of the time series. To mitigate the influence of the strength of the factors between rows and columns, we transformed the matrix factor model into a new one only including a row or column loading matrix. The improved ratio-based methods, based on this transformed model, were then investigated.
	
	We analyzed the theoretical properties of the proposed approaches under regularity conditions and demonstrated that they provide consistent estimators for the number of factors. Through Monte Carlo simulations and a real data analysis, we evaluated the finite-sample performance of the proposed methods and compared them with the traditional eigenvalue-based ratio approach. Our results show that the proposed estimators outperform the classical method, offering significant improvements in estimation accuracy.
	
	In summary, we believe that the methods introduced in this paper represent a valuable contribution to the toolbox for high-dimensional matrix-variate time series factor models, providing more reliable and accurate estimation techniques in practice. 
	
	\newpage
	\noindent {\Large \textbf{Appendix: Theorems, Lemmas and Proofs}}
	\setcounter{equation}{0}
	\renewcommand{\theequation}{A.\arabic{equation}}
	
	\section{Max-type and sum-type estimators}\label{sec:est}
	Assume that \(\bm{Q} = (Q_1, \dots, Q_r) = \bm{R}/p^{\frac{1-\delta}{2}}\) satisfies \(\bm{Q}^\top \bm{Q} = \bm{I}_r\), where \(Q_i \in \mathbb{R}^{p \times 1}\) for \(1 \leq i \leq r\). Let \(\bm{Q}^\dag = (Q_{r+1}, \dots, Q_p)\) be the orthogonal complement of \(\bm{Q}\), so that \((\bm{Q}, \bm{Q}^\dag)^\top (\bm{Q}, \bm{Q}^\dag) = \bm{I}_p\). To make our motivation clearer, we assume that $\bm E_t$ in model~(\ref{Model1}) is a white noise firstly. Thus, we have:
	\begin{eqnarray}
		Q_i^\top \bm{Y}_t &=& Q_i^\top \bm{R} \bm{F}_t \bm{C} + Q_i^\top \bm{E}_t, \quad i = 1, \dots, r; \label{Model2.1} \\
		Q_i^\top \bm{Y}_t &=& Q_i^\top \bm{E}_t, \quad i = r+1, \dots, p. \label{Model2.2}
	\end{eqnarray}
	Equations \eqref{Model2.1} and \eqref{Model2.2} indicate that \(Q_i^\top \bm{Y}_t\) is not a white noise sequence for \(i = 1, \dots, r\), while \(Q_i^\top \bm{Y}_t\) is a white noise sequence for \(i = r+1, \dots, p\). This motivates the development of a new approach to estimate the number of factors \(r\) by examining the uncorrelatedness of the sequences \(Q_i^\top \bm{Y}_t\) for \(i = 1, \dots, p\).
	
	Let \(\mathcal{R}_i = (Q_i, \dots, Q_p)\) for \(i = 1, \dots, p\). Our method relies on statistical tests to verify the white noise assumption for the sequence \(\{\mathcal{R}_i^\top \bm{Y}_t : t = 1, 2, \dots \}\). Specifically, we consider two types of test statistics: max-type statistics and sum-type statistics.
	
	For an integer \(h \geq 0\), we define the autocovariance matrix for \(\bm{Y}_t\) as:
	\begin{eqnarray*}
		\bm{\Sigma}_Y(h) &:=& \frac{1}{n-h} \sum_{t=1}^{n-h} \sum_{k=1}^{q} \text{Cov}(Y_{\cdot k,t}, Y_{\cdot k,t+h}) \\
		&=& \frac{1}{n-h} \sum_{t=1}^{n-h} \mathbb{E}(\bm{Y}_t \bm{Y}_{t+h}^\top) \in \mathbb{R}^{p \times p}.
	\end{eqnarray*}
	The auto-covariance matrix for \(\{\mathcal{R}_i^\top \bm{Y}_t\}\) at lag \(h\) is given by
	\begin{eqnarray} \label{Model5}
		\Gamma_i(h) := \Big( \gamma_{i,kl}(h) \Big) = \mathcal{R}_i^\top \bm{\Sigma}_Y(h) \mathcal{R}_i.
	\end{eqnarray}
	In our experience, using the estimator based on \eqref{Model5} provides better performance in determining the number of factors compared to using correlation matrices.
	The max-type test is constructed based on the maximum norm of the auto-covariance matrix:
	\begin{eqnarray} \label{Model6.1}
		T_{i,n} := \max_{1 \leq h \leq K} n^{1/2} \left| \Gamma_i(h) \right|_\infty = \max_{1 \leq h \leq K} n^{1/2} \left| \mathcal{R}_i^\top \bm{\Sigma}_Y(h) \mathcal{R}_i \right|_\infty.
	\end{eqnarray}
	Alternatively, the sum-type test is based on the Frobenius norm of the auto-covariance matrix:
	\begin{eqnarray} \label{Model6.2}
		G_{i,n} := \sum_{1 \leq h \leq K} \text{Tr} \left( \Gamma_i^\top(h) \Gamma_i(h) \right) = \sum_{1 \leq h \leq K} \left\| \Gamma_i(h) \right\|_F^2 = \sum_{1 \leq h \leq K} \sum_j \sigma_j^2 \left( \Gamma_i(h) \right),
	\end{eqnarray}
	where \(\sigma_j^2\) denotes the \(j\)-th largest singular value of \(\Gamma_{\mathcal{R}_i, Y}(h)\). Large values of \(T_{i,n}\) and \(G_{i,n}\) indicate a possible departure from the assumption of white noise in the sequence \(\{\mathcal{R}_i^\top \bm{Y}_t\}\). To estimate the number of factors \(r\), we sequentially check the values of these statistics as \(i\) increases. Intuitively, when \(i = r\), the gap between \(T_{i,n}\) and \(T_{i+1,n}\) is expected to be maximal, as \(\{\mathcal{R}_{r+1}^\top \bm{Y}_t\}\) is white noise, while \(\{\mathcal{R}_r^\top \bm{Y}_t\}\) is not. This insight forms the basis of our method, which is
    suitable for checking weakly correlated $\bm E_t$ and outlined in the following.
	
	\section{Estimation of the Number of Row Factors} \label{sec:est1}
	Throughout this section, we assume that \(\bm{Y}_t\) is a weakly stationary time series with zero mean. Under Assumption 2.1-(C5), we have the following expectations:
	\begin{eqnarray*}
		&&\mathbb{E}(\bm{R} \bm{F}_t \bm{C}^\top \bm{E}_{t+h}) =\text{trs}_p \left( \mathbb{E} \left( \text{vec}(\bm{R} \bm{F}_t) \text{vec}^\top (\bm{E}_{t+h} \bm{C}) \right) \right) = 0, \\
	&&\mathbb{E}(\bm{E}_t \bm{C} \bm{F}_t^\top \bm{R}^\top) = \text{trs}_p \left( \mathbb{E} \left( \text{vec} (\bm{E}_t \bm{C}) \text{vec}^\top (\bm{R} \bm{F}_{t+h}) \right) \right) = 0.
	\end{eqnarray*}
	Thus, for \(h > 0\), we have:
	\begin{eqnarray}
		\bm{\Sigma}_Y(h) &=& \mathbb{E}(\bm{Y}_t \bm{Y}_{t+h}^\top) = q^{1-\omega} \mathbb{E} \left( \bm{R} \bm{F}_t \bm{F}_{t+h}^\top \bm{R}^\top \right) + \mathbb{E} \left( \bm{E}_t \bm{C} \bm{F}_{t+h}^\top \bm{R}^\top \right) + \mathbb{E} (\bm{E}_t \bm{E}_{t+h}^\top) \nonumber \\
		&=& q^{1-\omega} \bm{R} \bm{\Sigma}_f(h) \bm{R}^\top + \text{Tr}_p \left( (\bm{C}^\top \otimes \bm{I}_p) \bm{\Sigma}_{ef}(h) (\bm{I}_c \otimes \bm{R}^\top) \right). \label{Model7}
	\end{eqnarray}
	
	For a pre-determined integer \(h_0 \geq 1\), we define:
	\begin{eqnarray} \label{Model70}
		\bm{M}_1 = \sum_{h=1}^{h_0} \bm{\Sigma}_Y^\top(h) \bm{\Sigma}_Y(h).
	\end{eqnarray}
	By Assumption 2.1, the matrix \(\bm{M}_1\) has rank \(r\), implying that every column of \(\bm{M}_1\) can be approximated by the first \(r\) eigenvectors of \(\bm{R}\). Therefore, the eigenspace of \(\bm{M}_1\) can be aligned with that of \(\bm{R}\), denoted as \(\mathcal{M}(\bm{R})\).
	It thus inspires us to estimate estimate \(\mathcal{M}(\bm{R})\) by the eigenvectors of the sample version of \(\bm{M}_1\), which is defined as:
	\begin{eqnarray*}
		\widehat{\bm{M}}_1 = \sum_{h=1}^{h_0} \widehat{\bm{\Sigma}}_Y^\top(h) \widehat{\bm{\Sigma}}_Y(h),
	\end{eqnarray*}
	where
	\[
	\widehat{\bm{\Sigma}}_Y(h) = \frac{1}{n} \sum_{t=1}^n \bm{Y}_t \bm{Y}_{t+h}^\top.
	\]
	
	Let \(\hat{Q}_i\) be the eigenvector of \(\widehat{\bm{M}}_1\) associated with its \(i\)-th largest eigenvalue for \(i = 1, 2, \dots, p\). Then, \(\mathcal{M}(\bm{R})\) can be estimated by \(\mathcal{M}(\widehat{\bm{Q}})\), where \(\widehat{\bm{Q}} = (\hat{Q}_1, \dots, \hat{Q}_r)\). Consequently, we estimate \(\bm{R}\) and \(\bm{Q}^\dag\) by \(\widehat{\bm{R}}_o = p^{\frac{1-\delta}{2}} \widehat{\bm{Q}}\) and \(\widehat{\bm{Q}}^\dag = (\hat{Q}_{r+1}, \dots, \hat{Q}_p)\), respectively.
	
	Based on these estimates, we define the feasible max-type and sum-type statistics as follows:
	\begin{align*}
		\widehat{T}_{i,n} &= \max_{1 \leq h \leq K} n^{1/2} \left| \widehat{\Gamma}_i(h) \right|_\infty, \\
		\widehat{G}_{i,n} &= \sum_{1 \leq h \leq K} \text{Tr} \left( \widehat{\Gamma}_i^\top(h) \widehat{\Gamma}_i(h) \right),
	\end{align*}
	where \(\widehat{\Gamma}_i(h) = \widehat{\mathcal{R}}_i^\top \widehat{\bm{\Sigma}}_Y(h) \widehat{\mathcal{R}}_i\).
	We propose the following ratio estimators for \(r\) based on an enhanced elbow criterion:
	\begin{eqnarray}
		\widehat{r}_{\text{MR}_o} &=& \arg \max_{1 \leq i \leq r_{\max}} \frac{\widehat{T}_{i,n}}{\widehat{T}_{i+1,n}} = \arg \max_{1 \leq i \leq r_{\max}} \text{MR}_o(i), \label{Model8.1} \\
		\widehat{r}_{\text{SR}_o} &=& \arg \max_{1 \leq i \leq r_{\max}} \frac{\widehat{G}_{i,n} - \widehat{G}_{i+1,n}}{\widehat{G}_{i+1,n} - \widehat{G}_{i+2,n}} = \arg \max_{1 \leq i \leq r_{\max}} \text{SR}_o(i), \label{Model8.2}
	\end{eqnarray}
	where \(\text{MR}_o(i) = \frac{\widehat{T}_{i,n}}{\widehat{T}_{i+1,n}}\) and \(\text{SR}_o(i) = \frac{\widehat{G}_{i,n} - \widehat{G}_{i+1,n}}{\widehat{G}_{i+1,n} - \widehat{G}_{i+2,n}}\). Here, \(r_{\max}\) is an upper bound specified by the user.
	The ratios based on max-type or sum-type statistics may exhibit significant variation, which could potentially affect the performance of these ratio estimators for \(r\).

	\subsection{Estimation of the number of column factors}\label{sec:est2}
	Define
	\begin{eqnarray} \label{Model70}
		&&\bm{M}_2 = \sum_{h=1}^{h_0} \widetilde{\bm{\Sigma}}_Y^\top(h) \widetilde{\bm{\Sigma}}_Y(h) \quad \text{and} \quad \widehat{\bm{M}}_2 = \sum_{h=1}^{h_0} \widehat{\widetilde{\bm{\Sigma}}}_Y^\top(h) \widehat{\widetilde{\bm{\Sigma}}}_Y(h),
	\end{eqnarray}
	where \(\widetilde{\bm{\Sigma}}_Y(h) = \frac{1}{n-h} \sum_{t=1}^{n-h} \mathbb{E}(\bm{Y}_t^\top \bm{Y}_{t+h})\) and its sample version is
	\(\widehat{\widetilde{\bm{\Sigma}}}_Y(h) = \frac{1}{n} \sum_{t=1}^{n-h} \bm{Y}_t^\top \bm{Y}_{t+h}\).
	
	Let \(\tilde{Q}_j\) and \(\hat{\tilde{Q}}_j\) be the unit eigenvectors corresponding to the \(j\)-th largest eigenvalues of \(\bm{M}_2\) and \(\widehat{\bm{M}}_2\), respectively. Denote \(\Gamma_j(h) = \mathcal{C}_j^\top \widetilde{\bm{\Sigma}}_Y(h) \mathcal{C}_j\), where \(\mathcal{C}_j = (\tilde{Q}_j, \dots, \tilde{Q}_q)\), for \(j = 1, \dots, q\).
	
	The max-type and sum-type statistics are defined as follows:
	\begin{eqnarray} \label{Model10.1}
		\widehat{\widetilde{T}}_{j,n} &=& \max_{1 \leq h \leq K} n^{1/2} \left|\widehat{\Gamma}_j(h)\right|_\infty,
	\end{eqnarray}
	and
	\begin{eqnarray} \label{Model10.2}
		\widehat{\widetilde{G}}_{j,n} &=& \sum_{1 \leq h \leq K} \text{Tr} \left( \widehat{\Gamma}_j^\top(h) \widehat{\Gamma}_j(h) \right),
	\end{eqnarray}
	where \(\widehat{\Gamma}_j(h) = \widehat{\mathcal{C}}_j^\top \widehat{\widetilde{\bm{\Sigma}}}_Y(h) \widehat{\mathcal{C}}_j\).
	
	The ratio estimators are defined as:
	\begin{eqnarray}
		&&\widehat{c}_{\mathrm{M R}_o} = \arg \max_{1 \leq j \leq c_{\max}} \frac{\widehat{\widetilde{T}}_{j,n}}{\widehat{\widetilde{T}}_{j+1,n}} =: \arg \max_{1 \leq j \leq c_{\max}} \widetilde{\mathrm{M R}}_o(j), \label{Model11.1} \\
		&&\widehat{c}_{\mathrm{S R}_o} = \arg \max_{1 \leq j \leq c_{\max}} \frac{\widehat{\widetilde{G}}_{j,n} - \widehat{\widetilde{G}}_{j+1,n}}{\widehat{\widetilde{G}}_{j+1,n} - \widehat{\widetilde{G}}_{j+2,n}} =: \arg \max_{1 \leq j \leq c_{\max}} \widetilde{\mathrm{S R}}_o(j), \label{Model11.2}
	\end{eqnarray}
	where \(c_{\max}\) is a user-defined upper bound.


	\subsection{Theoretical Properties}\label{sec:theory}
	\setcounter{equation}{0}
	\renewcommand{\theequation}{3.\arabic{equation}}
	In this section, we study the large sample properties of the proposed estimators in the asymptotic regime where \(p\), \(q\), and \(n\) tend to infinity while \(r\) and \(c\) are fixed. The following theorems establish the rates of convergence as well as the consistency for the ratios of the max-type and sum-type statistics.
	
	\begin{assumption}\label{assum-2}
		\textit{The matrix \(\bm M_1\) has \(r\) distinct positive eigenvalues, and the matrix \(\bm M_2\) has \(c\) distinct positive eigenvalues.}
	\end{assumption}
	
	\begin{theorem}\label{theo3}
		Under Assumptions~\ref{assum-1} and ~\ref{assum-2}, and the condition \(p^{\delta}q^{\omega}n^{-1/2} = \textbf{o}(1)\), we have
		\begin{eqnarray*}
			&&\dfrac{\widehat{T}_{r+1,n}}{\widehat{T}_{r,n}} = \textbf{O}_p\left(p^{2\delta}q^{2\omega}n^{-1}\right); \quad \dfrac{\widehat{\widetilde{T}}_{c+1,n}}{\widehat{\widetilde{T}}_{c,n}} = \textbf{O}_p\left(p^{2\delta}q^{2\omega}n^{-1}\right).
		\end{eqnarray*}
	\end{theorem}
	
	\begin{theorem}\label{theo2}
		Under Assumptions~\ref{assum-1} and ~\ref{assum-2}, and the condition \(p^{\delta}q^{\omega}n^{-1/2} = \textbf{o}(1)\), we have
		\begin{eqnarray*}
			&&\dfrac{\widehat{G}_{r+1,n} - \widehat{G}_{r+2,n}}{\widehat{G}_{r,n} - \widehat{G}_{r+1,n}} = \textbf{O}_p\left(p^{2\delta}q^{2\omega}n^{-1}\right); \quad \dfrac{\widehat{\widetilde{G}}_{c+1,n} - \widehat{\widetilde{G}}_{c+2,n}}{\widehat{\widetilde{G}}_{c,n} - \widehat{\widetilde{G}}_{c+1,n}} = \textbf{O}_p\left(p^{2\delta}q^{2\omega}n^{-1}\right).
		\end{eqnarray*}
	\end{theorem}
	
		\begin{theorem}\label{theo23} Suppose Assumptions~\ref{assum-1} and~\ref{assum-2} hold 
		and $p^{\delta}q^{\omega}n^{-1/2}=\textbf{o}(1)$. Then, the $\mathrm{MR}_o$, $\mathrm{SR}_o$, $\mathrm{\widetilde{MR}}_o$ and $\mathrm{\widetilde{SR}}_o$ estimators (i.e., $\widehat{r}_{\mathrm{MR}_o}$ in (\ref{Model8.1}), $\widehat{r}_{\mathrm{SR}_o}$ in (\ref{Model8.2}), $\widehat{c}_{\mathrm{M R}_o}$ in (\ref{Model11.1}) and $\widehat{c}_{\mathrm{S R}_o}$ in (\ref{Model11.2}))
        satisfy that
		\begin{eqnarray*}
			&&P(\widehat{r}_{\mathrm{SR}_o}=r)\to 1; \quad P(\widehat{c}_{\mathrm{SR}_o}=c)\to 1;\\
			&&P(\widehat{r}_{\mathrm{MR}_o}=r)\to 1; \quad P(\widehat{c}_{\mathrm{MR}_o}=c)\to 1.
		\end{eqnarray*}
	\end{theorem}

	\vskip 1 true cm
	
	\bibliographystyle{dcu}
	
	\bibliography{references}

	\begin{table*}
		\caption{Relative frequency estimates for $P(\hat{r} = 3)$ and $P(\hat{c} = 3)$
			with 200 reduplicate samples for the AR coefficients with case~(1) and $\bm\Sigma_1$ and $\bm\Sigma_2$ with case~(1) }
		\label{tab1}
		\scriptsize
		\begin{tabular*}{\textwidth}{@{\extracolsep{\fill}}ccccccccccc}
			\hline
			$(\delta,\omega)$& $(0.5,0.5)$ \\
			\hline
			$(p,q)$&$n$&$\quad ER_o$&$\quad SR_o$&$\quad MR_o$&$\quad \widetilde{ER}_o$&$\quad \widetilde{SR}_o$&$\quad \widetilde{MR}_o$\\
			\hline
			(20,20)& $200$ &0.360($128|0$)&0.640($70|2$)&0.415($114|3$)&0.340($130|2$)&0.640($70|2$)&0.370($124|2$)\\
			& $800$ &0.445($110|1$)&0.665($88|3$)&0.475($103|2$)&0.465($106|1$)&0.685($62|1$)&0.445($111|0$)\\
			\hline
			(20,40)& $200$ &0.355($170|1$)&0.480($103|1$)&0.345($131|0$)&0.470($106|0$)&0.910($18|0$)&0.580($83|1$)\\
			& $800$ &0.420($114|2$)&0.515($97|0$)&0.380($123|1$)&0.465($107|0$)&0.860($26|2$)&0.570($86|0$)\\
			\hline
			(40,40)& $200$ &0.430($113|1$)&0.770($46|0$)&0.535($92|1$)&0.390($122|0$)&0.790($42|0$)&0.565($87|0$)\\
			& $800$ &0.485($103|0$)&0.760($48|0$)&0.520($96|0$)&0.415($117|0$)&0.755($49|0$)&0.500($100|0$)\\
			\hline
			$(p,q)$&$n$&$\quad ER$&$\quad SR$&$\quad MR$&$\quad \widetilde{ER}$&$\quad \widetilde{SR}$&$\quad \widetilde{MR}$\\
			\hline
			(20,20)& $200$ &0.445($109|2$)&0.885($23|0$)&0.570($85|1$)&0.410($112|6$)&0.800($35|5$)&0.520($94|2$)\\
			& $800$ &0.545($88|3$)&0.875($21|4$)&0.560($84|4$)&0.495($96|5$)&0.895($19|2$)&0.630($71|2$)\\
			\hline
			(20,40)& $200$ &0.555($87|2$)&0.825($34|1$)&0.570($86|0$)&0.570($86|0$)&0.970($6|0$)&0.745($50|1$)\\
			& $800$ &0.545($91|0$)&0.855($28|1$)&0.595($81|0$)&0.545($88|3$)&0.955($5|4$)&0.715($56|1$)\\
			\hline
			(40,40)& $200$ &0.520($95|1$)&0.935($13|0$)&0.720($56|0$)&0.510($98|0$)&0.940($11|1$)&0.690($61|15$)\\
			& $800$ &0.575($85|0$)&0.970($5|1$)&0.705($59|0$)&0.485($102|1$)&0.965($7|0$)&0.690($62|0$)\\
			\hline
			$(\delta,\omega)$& $(0,0.5)$ \\
			\hline
			$(p,q)$&$n$&$\quad ER_o$&$\quad SR_o$&$\quad MR_o$&$\quad \widetilde{ER}_o$&$\quad \widetilde{SR}_o$&$\quad \widetilde{MR}_o$\\
			\hline
			(20,20)& $200$ &0.750($48|2$)&0.950($9|0$)&0.845($29|2$)&0.795($40|1$)&0.980($4|0$)&0.890($22|0$)\\
			& $800$ &0.770($44|2$)&0.965($7|0$)&0.875($25|0$)&0.805($38|1$)&0.985($3|0$)&0.875($24|1$)\\
			\hline
			(20,40)& $200$ &0.770($46|0$)&0.925($15|0$)&0.895($21|0$)&0.720($56|0$)&1.000($0|0$)&0.885($23|0$)\\
			& $800$ &0.750($50|0$)&0.920($16|0$)&0.870($26|0$)&0.735($53|0$)&1.000($0|0$)&0.915($17|0$)\\
			\hline
			(40,40)& $200$ &0.830($34|0$)&0.990($2|0$)&0.920($16|0$)&0.820($36|0$)&0.995($1|0$)&0.930($14|05$)\\
			& $800$ &0.840($32|0$)&0.990($2|0$)&0.935($13|0$)&0.820($36|0$)&0.990($2|0$)&0.920($16|0$)\\
			\hline
			$(p,q)$&$n$&$\quad ER$&$\quad SR$&$\quad MR$&$\quad \widetilde{ER}$&$\quad \widetilde{SR}$&$\quad \widetilde{MR}$\\
			\hline
			(20,20)& $200$ &0.755($45|4$)&0.980($2|2$)&0.895($17|4$)&0.800($39|1$)&1.000($0|0$)&0.910($16|2$)\\
			& $800$ &0.775($43|2$)&0.990($2|0$)&0.900($19|1$)&0.815($35|2$)&0.990($2|0$)&0.900($19|1$)\\
			\hline
			(20,40)& $200$ &0.790($42|0$)&0.980($4|0$)&0.890($22|0$)&0.715($57|0$)&1.000($0|0$)&0.905($17|2$)\\
			& $800$ &0.755($49|0$)&0.970($6|0$)&0.895($21|0$)&0.755($46|3$)&1.000($0|0$)&0.910($18|0$)\\
			\hline
			(40,40)& $200$ &0.830($34|0$)&1.000($0|0$)&0.960($16|0$)&0.825($35|0$)&1.000($0|0$)&0.930($13|1$)\\
			& $800$ &0.855($29|0$)&1.000($0|0$)&0.940($12|0$)&0.840($32|0$)&0.995($1|0$)&0.925($15|0$)\\
			\hline
			$(\delta,\omega)$& $(0,0)$ \\
			\hline
			$(p,q)$&$n$&$\quad ER_o$&$\quad SR_o$&$\quad MR_o$&$\quad \widetilde{ER}_o$&$\quad \widetilde{SR}_o$&$\quad \widetilde{MR}_o$\\
			\hline
			(20,20)& $200$ &0.910($18|0$)&1.000($0|0$)&0.985($3|0$)&0.910($18|0$)&1.000($0|0$)&0.980($2|0$)\\
			& $800$ &0.900($19|1$)&0.990($2|0$)&0.970($6|0$)&0.905($19|0$)&0.985($3|0$)&0.975($5|0$)\\
			\hline
			(20,40)& $200$ &0.965($7|0$)&1.000($0|0$)&0.985($3|0$)&0.910($18|0$)&1.000($0|0$)&0.970($6|0$)\\
			& $800$ &0.940($12|0$)&1.000($0|0$)&0.985($3|0$)&0.915($16|1$)&1.000($0|0$)&0.965($5|2$)\\
			\hline
			(40,40)& $200$ &0.925($15|0$)&1.000($0|0$)&0.965($6|1$)&0.955($9|0$)&1.000($0|0$)&0.985($3|0$)\\
			& $800$ &0.960($8|0$)&1.000($0|0$)&0.985($3|0$)&0.935($13|0$)&1.000($0|0$)&0.970($4|2$)\\
			\hline
			$(p,q)$&$n$&$\quad ER$&$\quad SR$&$\quad MR$&$\quad \widetilde{ER}$&$\quad \widetilde{SR}$&$\quad \widetilde{MR}$\\
			\hline
			(20,20)& $200$ &0.910($13|5$)&1.000($0|0$)&0.990($2|0$)&0.910($17|1$)&1.000($0|0$)&0.980($2|0$)\\
			& $800$ &0.900($18|1$)&0.990($2|0$)&0.970($5|1$)&0.900($18|2$)&0.995($1|0$)&0.975($5|0$)\\
			\hline
			(20,40)& $200$ &0.965($6|1$)&1.000($0|0$)&0.990($2|0$)&0.910($17|1$)&1.000($0|0$)&0.965($5|2$)\\
			& $800$ &0.945($10|1$)&1.000($0|0$)&0.995($1|0$)&0.915($15|2$)&1.000($0|0$)&0.955($5|4$)\\
			\hline
			(40,40)& $200$ &0.925($15|0$)&1.000($0|0$)&0.965($6|1$)&0.955($9|0$)&1.000($0|0$)&0.990($2|0$)\\
			& $800$ &0.960($7|1$)&1.000($0|0$)&0.995($1|0$)&0.935($13|0$)&1.000($0|0$)&0.965($3|4$)\\
			\hline
		\end{tabular*}
	\end{table*}
	
	\begin{table*}
		\caption{Relative frequency estimates for $P(\hat{r} = 3)$ and $P(\hat{c} = 3)$
			with 200 reduplicate samples for the AR coefficients with case~(2) and $\bm\Sigma_1$ and $\bm\Sigma_2$ with case~(1) }
		\label{tab2}
		\scriptsize
		\begin{tabular*}{\textwidth}{@{\extracolsep{\fill}}ccccccccccc}
			\hline
			$(\delta,\omega)$& $(0.5,0.5)$ \\
			\hline
			$(p,q)$&$n$&$\quad ER_o$&$\quad SR_o$&$\quad MR_o$&$\quad \widetilde{ER}_o$&$\quad \widetilde{SR}_o$&$\quad \widetilde{MR}_o$\\
			\hline
			(20,20)& $200$ &0.520($95|1$)&0.790($41|1$)&0.545($90|1$)&0.530($94|0$)&0.790($42|0$)&0.575($85|0$)\\
			& $800$ &0.500($99|1$)&0.775($45|0$)&0.525($95|0$)&0.460($106|2$)&0.775($44|1$)&0.515($96|1$)\\
			\hline
			(20,40)& $200$ &0.495($100|1$)&0.680($64|0$)&0.555($89|0$)&0.525($95|0$)&0.945($11|0$)&0.720($56|0$)\\
			& $800$ &0.535($92|1$)&0.710($58|0$)&0.550($89|1$)&0.525($90|5$)&0.955($9|0$)&0.775($44|1$)\\
			\hline
			(40,40)& $200$ &0.585($83|0$)&0.845($31|0$)&0.665($67|0$)&0.555($89|0$)&0.830($34|0$)&0.640($72|0$)\\
			& $800$ &0.530($94|0$)&0.905($10|0$)&0.650($70|0$)&0.545($91|0$)&0.885($23|0$)&0.620($76|0$)\\
			\hline
			$(p,q)$&$n$&$\quad ER$&$\quad SR$&$\quad MR$&$\quad \widetilde{ER}$&$\quad \widetilde{SR}$&$\quad \widetilde{MR}$\\
			\hline
			(20,20)& $200$ &0.585($79|4$)&0.905($17|2$)&0.685($63|0$)&0.565($83|4$)&0.895($20|1$)&0.700($60|0$)\\
			& $800$ &0.560($82|6$)&0.920($13|3$)&0.630($72|2$)&0.530($92|2$)&0.905($16|3$)&0.690($59|3$)\\
			\hline
			(20,40)& $200$ &0.600($79|1$)&0.850($30|0$)&0.690($62|0$)&0.565($87|0$)&0.975($4|1$)&0.815($37|0$)\\
			& $800$ &0.625($73|2$)&0.900($20|0$)&0.710($57|1$)&0.675($62|3$)&0.970($6|0$)&0.850($29|1$)\\
			\hline
			(40,40)& $200$ &0.655($69|0$)&0.955($8|1$)&0.780($44|0$)&0.605($79|0$)&0.925($15|0$)&0.785($42|1$)\\
			& $800$ &0.590($82|0$)&0.975($4|1$)&0.810($38|0$)&0.615($77|0$)&0.965($7|0$)&0.740($51|1$)\\
			\hline
			$(\delta,\omega)$& $(0,0.5)$ \\
			\hline
			$(p,q)$&$n$&$\quad ER_o$&$\quad SR_o$&$\quad MR_o$&$\quad \widetilde{ER}_o$&$\quad \widetilde{SR}_o$&$\quad \widetilde{MR}_o$\\
			\hline
			(20,20)& $200$ &0.850($30|0$)&0.965($7|0$)&0.890($22|0$)&0.830($33|1$)&0.985($3|0$)&0.885($22|1$)\\
			& $800$ &0.800($40|0$)&0.975($5|0$)&0.820($26|0$)&0.815($37|0$)&0.990($2|0$)&0.935($13|0$)\\
			\hline
			(20,40)& $200$ &0.855($28|1$)&0.960($8|0$)&0.900($20|0$)&0.815($37|0$)&1.000($0|0$)&0.920($12|0$)\\
			& $800$ & 0.830($34|0$)&0.920($16|0$)&0.925($15|0$)&0.815($37|0$)&1.000($0|0$)&0.920($12|0$)\\
			\hline
			(40,40)& $200$ &0.815($37|0$)&0.995($1|0$)&0.940($12|0$)&0.840($32|0$)&1.000($0|0$)&0.940($12|0$)\\
			& $800$ &0.845($31|0$)&0.995($1|0$)&0.950($10|0$)&0.875($25|0$)&0.985($3|0$)&0.945($11|0$)\\
			\hline
			$(p,q)$&$n$&$\quad ER$&$\quad SR$&$\quad MR$&$\quad \widetilde{ER}$&$\quad \widetilde{SR}$&$\quad \widetilde{MR}$\\
			\hline
			(20,20)& $200$ &0.850($26|4$)&0.995($1|0$)&0.925($15|0$)&0.835($31|2$)&1.000($0|0$)&0.900($18|2$)\\
			& $800$ &0.810($36|2$)&0.985($3|0$)&0.895($20|1$)&0.825($34|1$)&0.995($0|1$)&0.940($11|1$)\\
			\hline
			(20,40)& $200$&0.865($26|1$)&0.980($3|1$)&0.925($15|0$)&0.825($34|2$)&1.000($0|0$)&0.910($12|2$)\\
			& $800$ & 0.830($34|0$)&0.980($4|0$)&0.955($9|0$)&0.825($32|3$)&1.000($0|0$)&0.915($12|1$)\\
			\hline
			(40,40)& $200$ &0.840($32|0$)&0.995($1|0$)&0.945($10|1$)&0.850($30|0$)&1.000($0|0$)&0.950($9|1$)\\
			& $800$ &0.850($30|0$)&1.000($0|0$)&0.945($11|0$)&0.880($24|0$)&1.000($0|0$)&0.945($11|0$)\\
			\hline
			$(\delta,\omega)$& $(0,0)$ \\
			\hline
			$(p,q)$&$n$&$\quad ER_o$&$\quad SR_o$&$\quad MR_o$&$\quad \widetilde{ER}_o$&$\quad \widetilde{SR}_o$&$\quad \widetilde{MR}_o$\\
			\hline
			(20,20)& $200$ &0.930($12|0$)&1.000($0|0$)&0.990($2|0$)&0.910($16|2$)&1.000($0|0$)&0.965($5|2$)\\
			& $800$ &0.930($12|2$)&1.000($0|0$)&0.990($2|0$)&0.935($13|0$)&1.000($0|0$)&0.975($5|0$)\\
			\hline
			(20,40)& $200$ &0.975($5|0$)&1.000($0|0$)&1.000($0|0$)&0.945($11|0$)&1.000($0|0$)&0.975($5|0$)\\
			& $800$ &0.970($6|0$)&1.000($0|0$)&1.000($0|0$)&0.940($10|0$)&1.000($0|0$)&0.805($39|0$)\\
			\hline
			(40,40)& $200$ &0.960($8|0$)&1.000($0|0$)&1.000($0|0$)&0.980($3|1$)&1.000($0|0$)&0.995($10$)\\
			& $800$ &0.970($6|0$)&0.995($1|0$)&0.770($46|0$)&0.965($7|0$)&1.000($0|0$)&0.980($2|2$)\\
			\hline
			$(p,q)$&$n$&$\quad ER$&$\quad SR$&$\quad MR$&$\quad \widetilde{ER}$&$\quad \widetilde{SR}$&$\quad \widetilde{MR}$\\
			\hline
			(20,20)& $200$ &0.930($10|2$)&1.000($0|0$)&0.995($0|1$)&0.905($12|7$)&1.000($0|0$)&0.970($4|2$)\\
			& $800$ &0.930($10|4$)&1.000($0|0$)&0.990($2|0$)&0.930($12|2$)&1.000($0|0$)&0.975($5|0$)\\
			\hline
			(20,40)& $200$ &0.975($5|0$)&1.000($0|0$)&1.000($0|0$)&0.940($11|1$)&1.000($0|0$)&0.975($4|1$)\\
			& $800$ &0.970($6|0$)&1.000($0|0$)&1.000($0|0$)&0.945($9|2$)&1.000($0|0$)&0.980($1|3$)\\
			\hline
			(40,40)& $200$ &0.960($8|0$)&1.000($0|0$)&1.000($0|0$)&0.980($3|1$)&1.000($0|0$)&0.990($0|2$)\\
			& $800$ &0.970($6|0$)&1.000($0|0$)&1.000($0|0$)&0.965($6|1$)&1.000($0|0$)&0.985($1|2$)\\
			\hline
		\end{tabular*}
	\end{table*}
	\begin{table*}
		\caption{Relative frequency estimates for $P(\hat{r} = 3)$ and $P(\hat{c} = 3)$
			with 200 reduplicate samples for the AR coefficients with case~(3) and $\bm\Sigma_1$ and $\bm\Sigma_2$ with case~(1) }
		\label{tab3}
		\scriptsize
		\begin{tabular*}{\textwidth}{@{\extracolsep{\fill}}ccccccccccc}
			\hline
			$(\delta,\omega)$& $(0.5,0.5)$ \\
			\hline
			$(p,q)$&$n$&$\quad ER_o$&$\quad SR_o$&$\quad MR_o$&$\quad \widetilde{ER}_o$&$\quad \widetilde{SR}_o$&$\quad \widetilde{MR}_o$\\
			\hline
			(20,20)& $200$ &0.825($35|0$)&0.960($8|0$)&0.915($17|0$)&0.795($41|0$)&0.960($8|0$)&0.925($15|0$)\\
			& $800$ &0.820($36|0$)&0.985($3|0$)&0.960($8|0$)&0.850($30|0$)&0.995($1|0$)&0.935($13|0$)\\
			\hline
			(20,40)& $200$ &0.840($32|0$)&0.965($7|0$)&0.950($10|0$)&0.800($40|0$)&0.990($2|0$)&0.935($13|0$)\\
			& $800$ &0.840($32|0$)&0.975($5|0$)&0.955($9|0$)&0.805($39|0$)&0.990($2|0$)&0.945($11|0$)\\
			\hline
			(40,40)& $200$ &0.810($38|0$)&0.980($4|0$)&0.970($6|0$)&0.855($29|0$)&0.975($5|0$)&0.940($12|05$)\\
			& $800$ &0.855($29|0$)&0.990($2|0$)&0.965($7|0$)&0.805($39|0$)&0.990($2|0$)&0.935($13|0$)\\
			\hline
			$(p,q)$&$n$&$\quad ER$&$\quad SR$&$\quad MR$&$\quad \widetilde{ER}$&$\quad \widetilde{SR}$&$\quad \widetilde{MR}$\\
			\hline
			(20,20)& $200$ &0.825($32|3$)&1.000($0|0$)&0.940($11|1$)&0.810($37|1$)&0.980($4|0$)&0.955($9|0$)\\
			& $800$ &0.835($33|0$)&1.000($3|0$)&0.975($5|0$)&0.860($26|2$)&0.995($1|0$)&0.960($8|0$)\\
			\hline
			(20,40)& $200$ &0.840($32|0$)&0.985($3|0$)&0.970($6|0$)&0.800($39|1$)&0.990($2|0$)&0.930($13|1$)\\
			& $800$ &0.845($31|0$)&0.995($1|0$)&0.975($5|0$)&0.810($38|0$)&0.995($0|1$)&0.940($12|0$)\\
			\hline
			(40,40)& $200$ &0.825($35|0$)&0.985($3|0$)&0.975($5|0$)&0.865($27|0$)&0.990($2|0$)&0.940($12|05$)\\
			& $800$ &0.870($26|0$)&1.000($0|0$)&0.965($6|1$)&0.825($34|1$)&0.990($2|0$)&0.935($12|1$)\\
			\hline
			$(\delta,\omega)$& $(0,0.5)$ \\
			\hline
			$(p,q)$&$n$&$\quad ER_o$&$\quad SR_o$&$\quad MR_o$&$\quad \widetilde{ER}_o$&$\quad \widetilde{SR}_o$&$\quad \widetilde{MR}_o$\\
			\hline
			(20,20)& $200$ &0.940($12|0$)&1.000($0|0$)&0.985($3|0$)&0.915($17|0$)&1.000($0|0$)&0.995($1|0$)\\
			& $800$ &0.960($8|0$)&0.990($2|0$)&0.980($3|1$)&0.940($12|0$)&1.000($0|0$)&0.990($2|0$)\\
			\hline
			(20,40)& $200$ &0.955($9|0$)&0.985($3|0$)&0.970($6|0$)&0.915($17|0$)&1.000($0|0$)&0.970($4|2$)\\
			& $800$ &0.955($9|0$)&0.985($3|0$)&0.990($2|0$)&0.950($10|0$)&1.000($0|0$)&1.000($0|0$)\\
			\hline
			(40,40)& $200$ &0.985($3|0$)&1.000($0|0$)&0.995($1|0$)&0.970($6|0$)&1.000($0|0$)&0.985($1|2$)\\
			& $800$ &0.970($6|0$)&1.000($0|0$)&0.990($1|1$)&0.955($9|0$)&1.000($0|0$)&0.990($2|0$)\\
			\hline
			$(p,q)$&$n$&$\quad ER$&$\quad SR$&$\quad MR$&$\quad \widetilde{ER}$&$\quad \widetilde{SR}$&$\quad \widetilde{MR}$\\
			\hline
			(20,20)& $200$ &0.935($11|2$)&1.000($0|0$)&0.985($2|1$)&0.915($16|1$)&1.000($0|0$)&0.990($1|1$)\\
			& $800$ &0.950($7|3$)&0.995($1|0$)&0.980($2|2$)&0.930($11|3$)&1.000($0|0$)&0.990($2|0$)\\
			\hline
			(20,40)& $200$ &0.960($7|1$)&0.990($3|0$)&0.975($5|0$)&0.910($16|2$)&1.000($0|0$)&0.965($5|2$)\\
			& $800$ &0.965($7|0$)&0.990($2|0$)&0.990($1|1$)&0.950($10|0$)&1.000($0|0$)&1.000($0|0$)\\
			\hline
			(40,40)& $200$ &0.985($3|0$)&1.000($0|0$)&0.995($1|0$)&0.970($6|0$)&1.000($0|0$)&0.980($3|15$)\\
			& $800$ &0.975($5|0$)&1.000($0|0$)&0.995($0|0$)&0.965($7|0$)&1.000($0|0$)&0.985($1|2$)\\
			\hline
			$(\delta,\omega)$& $(0,0)$ \\
			\hline
			$(p,q)$&$n$&$\quad ER_o$&$\quad SR_o$&$\quad MR_o$&$\quad \widetilde{ER}_o$&$\quad \widetilde{SR}_o$&$\quad \widetilde{MR}_o$\\
			\hline
			(20,20)& $200$ &0.985($2|1$)&1.000($0|0$)&1.000($1|0$)&0.980($4|0$)&0.995($1|0$)&0.985($2|1$)\\
			& $800$ &0.990($2|0$)&0.995($1|0$)&0.995($1|0$)&0.990($1|3$)&1.000($0|0$)&1.000($0|0$)\\
			\hline
			(20,40)& $200$  &0.985($2|1$)&1.000($0|0$)&0.995($1|0$)&0.980($4|0$)&1.000($0|0$)&1.000($0|0$)\\
			& $800$ &0.990($2|0$)&1.000($1|0$)&0.995($1|0$)&0.990($2|0$)&1.000($0|0$)&1.000($0|0$)\\
			\hline
			(40,40)& $200$ &0.980($3|1$)&1.000($0|0$)&1.000($0|0$)&0.990($2|0$)&1.000($0|0$)&1.000($0|0$)\\
			& $800$ &1.000($0|0$)&1.000($0|0$)&1.000($0|0$)&1.000($0|0$)&1.000($0|0$)&1.000($0|0$)\\
			\hline
			$(p,q)$&$n$&$\quad ER$&$\quad SR$&$\quad MR$&$\quad \widetilde{ER}$&$\quad \widetilde{SR}$&$\quad \widetilde{MR}$\\
			\hline
			(20,20)& $200$ &0.980($2|2$)&1.000($0|0$)&1.000($0|0$)&0.980($4|0$)&0.995($1|0$)&0.985($2|1$)\\
			& $800$ &0.985($2|3$)&0.995($1|0$)&0.995($1|0$)&0.980($1|3$)&1.000($0|0$)&1.000($0|0$)\\
			\hline
			(20,40)& $200$ &0.985($2|1$)&1.000($0|0$)&1.000($0|0$)&0.970($4|2$)&1.000($0|0$)&1.000($0|0$)\\
			& $800$ &0.985($1|2$)&1.000($0|0$)&0.995($1|0$)&0.985($2|1$)&1.000($0|0$)&1.000($0|0$)\\
			\hline
			(40,40)& $200$ &0.975($3|2$)&0.975($5|0$)&0.790($42|0$)&0.990($1|1$)&1.000($0|0$)&1.000($0|0$)\\
			& $800$ &1.000($0|0$)&1.000($0|0$)&1.000($0|0$)&1.000($0|0$)&1.000($0|0$)&1.000($0|0$)\\
			\hline
		\end{tabular*}
	\end{table*}
	
	\begin{table*}
		\caption{Relative frequency estimates for $P(\hat{r} = 3)$ and $P(\hat{c} = 3)$
			with 200 reduplicate samples for the AR coefficients with case~(1) and $\bm\Sigma_1$ and $\bm\Sigma_2$ with case~(2) }
		\label{tab4}
		\scriptsize
		\begin{tabular*}{\textwidth}{@{\extracolsep{\fill}}ccccccccccc}
			\hline
			$(\delta,\omega)$& $(0.5,0.5)$ \\
			\hline
			$(p,q)$&$n$&$\quad ER_o$&$\quad SR_o$&$\quad MR_o$&$\quad \widetilde{ER}_o$&$\quad \widetilde{SR}_o$&$\quad \widetilde{MR}_o$\\
			\hline
			(20,20)& $200$ &0.335($132|1$)&0.540($85|7$)&0.330($130|4$)&0.360($125|3$)&0.540($85|7$)&0.360($127|1$)\\
			& $800$ &0.330($132|2$)&0.660($81|7$)&0.365($127|0$)&0.395($120|1$)&0.635($69|4$)&0.465($106|1$)\\
			\hline
			(20,40)& $200$ &0.335($131|2$)&0.395($118|3$)&0.375($124|1$)&0.375($125|0$)&0.600($68|12$)&0.455($106|3$)\\
			& $800$ &0.365($126|1$)&0.370($122|4$)&0.365($124|3$)&0.435($111|2$)&0.660($51|17$)&0.420($111|5$)\\
			\hline
			(40,40)& $200$ &0.360($125|3$)&0.310($118|20$)&0.315($119|20$)&0.345($129|2$)&0.325($103|32$)&0.325($132|3$)\\
			& $800$ &0.335($129|4$)&0.370($107|19$)&0.410($111|7$)&0.345($129|2$)&0.330($107|27$)&0.330($123|11$)\\
			\hline
			$(p,q)$&$n$&$\quad ER$&$\quad SR$&$\quad MR$&$\quad \widetilde{ER}$&$\quad \widetilde{SR}$&$\quad \widetilde{MR}$\\
			\hline
			(20,20)& $200$ &0.425($111|4$)&0.820($32|4$)&0.535($89|4$)&0.480($101|3$)&0.790($36|6$)&0.570($83|3$)\\
			& $800$ &0.440($108|4$)&0.825($31|4$)&0.550($88|2$)&0.510($97|1$)&0.855($28|1$)&0.550($82|8$)\\
			\hline
			(20,40)& $200$ &0.445($109|2$)&0.690($60|2$)&0.510($96|2$)&0.520($95|1$)&0.890($18|4$)&0.670($65|1$)\\
			& $800$ &0.460($108|0$)&0.655($69|0$)&0.490($101|0$)&0.565($86|1$)&0.930($9|5$)&0.675($65|0$)\\
			\hline
			(40,40)& $200$ &0.470($102|4$)&0.635($59|14$)&0.480($102|2$)&0.510($98|0$)&0.680($51|13$)&0.510($95|3$)\\
			& $800$ &0.510($96|2$)&0.740($44|8$)&0.620($71|5$)&0.455($126|3$)&0.760($41|7$)&0.500($95|5$)\\
			\hline
			$(\delta,\omega)$& $(0,0.5)$ \\
			\hline
			$(p,q)$&$n$&$\quad ER_o$&$\quad SR_o$&$\quad MR_o$&$\quad \widetilde{ER}_o$&$\quad \widetilde{SR}_o$&$\quad \widetilde{MR}_o$\\
			\hline
			(20,20)& $200$ &0.725($55|0$)&0.925($14|1$)&0.835($32|1$)&0.735($53|0$)&0.955($9|0$)&0.855($29|0$)\\
			& $800$ &0.785($43|0$)&0.965($6|1$)&0.885($23|0$)&0.765($47|0$)&0.945($11|0$)&0.835($33|0$)\\
			\hline
			(20,40)& $200$ &0.820($36|0$)&0.930($14|0$)&0.840($32|0$)&0.725($55|0$)&0.995($0|1$)&0.845($31|0$)\\
			& $800$ &0.835($33|0$)&0.960($8|0$)&0.885($23|0$)&0.710($58|0$)&0.970($6|0$)&0.885($23|0$)\\
			\hline
			(40,40)& $200$ &0.815($37|0$)&0.975($5|0$)&0.915($16|1$)&0.825($35|0$)&0.985($3|0$)&0.895($21|0$)\\
			& $800$ &0.885($23|0$)&0.990($0|0$)&0.890($22|0$)&0.820($35|0$)&0.985($3|0$)&0.925($15|0$)\\
			\hline
			$(p,q)$&$n$&$\quad ER$&$\quad SR$&$\quad MR$&$\quad \widetilde{ER}$&$\quad \widetilde{SR}$&$\quad \widetilde{MR}$\\
			\hline
			(20,20)& $200$ &0.725($53|2$)&0.960($7|1$)&0.840($28|4$)&0.740($52|0$)&0.990($2|0$)&0.910($18|0$)\\
			& $800$ &0.785($41|2$)&0.990($1|1$)&0.915($15|2$)&0.785($41|2$)&0.955($9|0$)&0.880($24|0$)\\
			\hline
			(20,40)& $200$ &0.830($34|0$)&0.960($8|0$)&0.900($20|0$)&0.730($53|1$)&0.995($1|0$)&0.885($22|1$)\\
			& $800$ &0.835($33|0$)&0.980($4|0$)&0.910($17|1$)&0.725($53|2$)&0.985($2|1$)&0.890($21|1$)\\
			\hline
			(40,40)& $200$ &0.835($33|0$)&1.000($0|0$)&0.915($15|2$)&0.830($33|1$)&0.995($1|0$)&0.905($19|0$)\\
			& $800$ &0.880($21|3$)&1.000($0|0$)&0.905($19|0$)&0.820($35|0$)&1.000($0|0$)&0.940($12|0$)\\
			\hline
			$(\delta,\omega)$& $(0,0)$ \\
			\hline
			$(p,q)$&$n$&$\quad ER_o$&$\quad SR_o$&$\quad MR_o$&$\quad \widetilde{ER}_o$&$\quad \widetilde{SR}_o$&$\quad \widetilde{MR}_o$\\
			\hline
			(20,20)& $200$ &0.890($21|1$)&0.995($1|0$)&0.945($11|0$)&0.910($16|2$)&0.980($4|0$)&0.930($14|0$)\\
			& $800$ &0.925($14|1$)&0.990($2|0$)&0.960($6|2$)&0.920($15|1$)&0.995($1|0$)&0.980($4|0$)\\
			\hline
			(20,40)& $200$  &0.970($6|0$)&0.990($2|0$)&0.990($2|0$)&0.950($10|0$)&1.000($0|0$)&0.965($4|3$)\\
			& $800$ &0.935($13|0$)&0.985($3|0$)&0.980($4|0$)&0.935($12|1$)&1.000($0|0$)&0.955($7|2$)\\
			\hline
			(40,40)& $200$ &0.980($4|0$)&0.995($1|0$)&0.990($2|0$)&0.945($11|0$)&1.000($0|0$)&0.990($0|2$)\\
			& $800$ &0.960($8|0$)&0.995($1|0$)&0.995($1|0$)&0.940($12|0$)&1.000($0|0$)&0.995($1|0$)\\
			\hline
			$(p,q)$&$n$&$\quad ER$&$\quad SR$&$\quad MR$&$\quad \widetilde{ER}$&$\quad \widetilde{SR}$&$\quad \widetilde{MR}$\\
			\hline
			(20,20)& $200$ &0.890($19|3$)&1.000($0|0$)&0.970($6|0$)&0.910($15|3$)&0.990($2|0$)&0.950($10|0$)\\
			& $800$ &0.925($13|2$)&0.995($1|0$)&0.995($1|0$)&0.920($14|2$)&0.995($1|0$)&0.985($3|0$)\\
			\hline
			(20,40)& $200$ &0.970($6|0$)&0.995($1|0$)&0.995($1|0$)&0.950($9|1$)&1.000($0|0$)&0.960($5|3$)\\
			& $800$ &0.940($11|1$)&0.995($1|0$)&0.980($4|0$)&0.930($11|3$)&1.000($0|0$)&0.960($6|2$)\\
			\hline
			(40,40)& $200$ &0.975($3|2$)&0.995($1|0$)&0.990($2|0$)&0.945($11|0$)&1.000($0|0$)&0.990($0|2$)\\
			& $800$ &0.960($8|0$)&0.995($1|0$)&0.995($1|0$)&0.940($12|0$)&1.000($0|0$)&0.995($1|0$)\\
			\hline
		\end{tabular*}
	\end{table*}
	
	\begin{table*}
		\caption{Relative frequency estimates for $P(\hat{r} = 3)$ and $P(\hat{c} = 3)$
			with 200 reduplicate samples for the AR coefficients with case~(2) and $\bm\Sigma_1$ and $\bm\Sigma_2$ with case~(2) }
		\label{tab5}
		\scriptsize
		\begin{tabular*}{\textwidth}{@{\extracolsep{\fill}}ccccccccccc}
			\hline
			$(\delta,\omega)$& $(0.5,0.5)$ \\
			\hline
			$(p,q)$&$n$&$\quad ER_o$&$\quad SR_o$&$\quad MR_o$&$\quad \widetilde{ER}_o$&$\quad \widetilde{SR}_o$&$\quad \widetilde{MR}_o$\\
			\hline
			(20,20)& $200$ &0.500($97|3$)&0.670($66|0$)&0.535($91|2$)&0.400($118|2$)&0.680($63|1$)&0.540($91|1$)\\
			& $800$ &0.525($95|0$)&0.690($55|7$)&0.500($97|3$)&0.515($96|1$)&0.725($54|1$)&0.510($97|1$)\\
			\hline
			(20,40)& $200$ &0.415($114|3$)&0.540($90|2$)&0.510($98|0$)&0.510($98|0$)&0.830($31|3$)&0.630($73|1$)\\
			& $800$ &0.470($106|0$)&0.535($92|1$)&0.485($103|0$)&0.550($89|1$)&0.770($39|7$)&0.570($84|2$)\\
			\hline
			(40,40)& $200$ &0.490($102|0$)&0.490($96|6$)&0.460($105|3$)&0.480($104|0$)&0.545($88|3$)&0.555($87|2$)\\
			& $800$ &0.500($100|0$)&0.565($78|9$)&0.585($81|2$)&0.450($110|0$)&0.485($91|12$)&0.500($98|2$)\\
			\hline
			$(p,q)$&$n$&$\quad ER$&$\quad SR$&$\quad MR$&$\quad \widetilde{ER}$&$\quad \widetilde{SR}$&$\quad \widetilde{MR}$\\
			\hline
			(20,20)& $200$ &0.555($85|4$)&0.870($26|0$)&0.620($76|0$)&0.535($91|2$)&0.905($18|1$)&0.635($69|4$)\\
			& $800$ &0.530($91|3$)&0.840($29|3$)&0.675($63|2$)&0.585($81|2$)&0.880($22|2$)&0.650($67|3$)\\
			\hline
			(20,40)& $200$ &0.500($99|0$)&0.750($49|1$)&0.615($76|1$)&0.620($74|2$)&0.985($1|2$)&0.790($41|1$)\\
			& $800$ &0.610($78|0$)&0.795($41|0$)&0.610($78|0$)&0.675($65|0$)&0.960($0|1$)&0.755($47|2$)\\
			\hline
			(40,40)& $200$ &0.550($90|0$)&0.830($31|3$)&0.660($68|0$)&0.575($85|0$)&0.840($30|2$)&0.705($58|1$)\\
			& $800$ &0.655($69|0$)&0.905($19|0$)&0.735($51|2$)&0.565($86|1$)&0.875($23|2$)&0.670($65|1$)\\
			\hline
			$(\delta,\omega)$& $(0,0.5)$ \\
			\hline
			$(p,q)$&$n$&$\quad ER_o$&$\quad SR_o$&$\quad MR_o$&$\quad \widetilde{ER}_o$&$\quad \widetilde{SR}_o$&$\quad \widetilde{MR}_o$\\
			\hline
			(20,20)& $200$ &0.790($41|1$)&0.975($5|0$)&0.895($21|0$)&0.815($37|0$)&0.985($3|0$)&0.915($17|0$)\\
			& $800$ &0.795($41|0$)&0.970($5|1$)&0.890($21|1$)&0.840($30|0$)&0.950($10|0$)&0.910($18|0$)\\
			\hline
			(20,40)& $200$ &0.880($24|0$)&0.940($12|0$)&0.895($21|0$)&0.785($42|1$)&0.995($0|1$)&0.915($15|2$)\\
			& $800$ &0.830($34|0$)&0.955($9|0$)&0.915($17|0$)&0.790($42|0$)&0.990($2|0$)&0.895($21|0$)\\
			\hline
			(40,40)& $200$ &0.855($29|0$)&1.000($0|0$)&0.995($1|0$)&0.850($30|0$)&0.990($2|0$)&0.910($18|0$)\\
			& $800$ &0.890($22|0$)&0.980($4|0$)&0.945($11|0$)&0.855($29|0$)&1.000($0|0$)&0.925($15|0$)\\
			\hline
			$(p,q)$&$n$&$\quad ER$&$\quad SR$&$\quad MR$&$\quad \widetilde{ER}$&$\quad \widetilde{SR}$&$\quad \widetilde{MR}$\\
			\hline
			(20,20)& $200$ &0.785($42|1$)&0.995($1|0$)&0.900($18|2$)&0.830($34|0$)&0.995($1|0$)&0.930($13|1$)\\
			& $800$ &0.795($41|0$)&0.985($3|0$)&0.910($17|1$)&0.845($28|3$)&0.975($5|0$)&0.925($15|0$)\\
			\hline
			(20,40)& $200$ &0.885($23|0$)&0.970($6|0$)&0.910($18|0$)&0.780($43|1$)&0.995($0|1$)&0.925($13|2$)\\
			& $800$ &0.850($30|0$)&0.975($5|0$)&0.935($13|0$)&0.810($37|1$)&0.995($1|0$)&0.910($18|0$)\\
			\hline
			(40,40)& $200$ &0.855($28|1$)&1.000($0|0$)&0.995($1|0$)&0.850($30|0$)&0.995($1|0$)&0.925($15|0$)\\
			& $800$ &0.900($20|0$)&1.000($0|0$)&0.950($10|0$)&0.870($26|0$)&1.000($0|0$)&0.920($14|2$)\\
			\hline
			$(\delta,\omega)$& $(0,0)$ \\
			\hline
			$(p,q)$&$n$&$\quad ER_o$&$\quad SR_o$&$\quad MR_o$&$\quad \widetilde{ER}_o$&$\quad \widetilde{SR}_o$&$\quad \widetilde{MR}_o$\\
			\hline
			(20,20)& $200$ &0.935($13|0$)&1.000($0|0$)&0.970($5|1$)&0.920($16|0$)&0.990($2|0$)&0.970($5|1$)\\
			& $800$ &0.905($18|1$)&1.000($0|0$)&0.985($2|1$)&0.950($8|2$)&0.990($2|0$)&0.980($4|0$)\\
			\hline
			(20,40)& $200$  &0.970($6|0$)&1.000($0|0$)&0.995($1|0$)&0.950($8|2$)&1.000($0|0$)&0.990($1|1$)\\
			& $800$ &0.975($5|0$)&0.995($1|0$)&0.995($1|0$)&0.950($9|1$)&1.000($0|0$)&0.990($2|0$)\\
			\hline
			(40,40)& $200$ &0.955($9|0$)&1.000($0|0$)&0.980($2|0$)&0.970($6|0$)&1.000($0|0$)&0.995($1|0$)\\
			& $800$ &0.965($6|0$)&1.000($0|0$)&1.000($0|0$)&0.960($8|0$)&1.000($0|0$)&0.990($1|1$)\\
			\hline
			$(p,q)$&$n$&$\quad ER$&$\quad SR$&$\quad MR$&$\quad \widetilde{ER}$&$\quad \widetilde{SR}$&$\quad \widetilde{MR}$\\
			\hline
			(20,20)& $200$ &0.930($11|3$)&1.000($0|0$)&0.970($5|1$)&0.920($14|2$)&1.000($0|0$)&0.970($5|1$)\\
			& $800$ &0.905($17|2$)&1.000($0|0$)&0.985($2|1$)&0.950($7|3$)&1.000($0|0$)&0.980($4|0$)\\
			\hline
			(20,40)& $200$ &0.970($5|1$)&1.000($0|0$)&0.995($1|0$)&0.950($8|2$)&1.000($0|0$)&0.995($0|1$)\\
			& $800$ &0.975($5|0$)&0.995($1|0$)&0.995($1|0$)&0.950($8|2$)&1.000($0|0$)&0.990($2|0$)\\
			\hline
			(40,40)& $200$ &0.950($9|0$)&1.000($0|0$)&0.985($1|2$)&0.970($6|0$)&1.000($0|0$)&0.995($1|0$)\\
			& $800$ &0.965($6|0$)&1.000($0|0$)&1.000($0|0$)&0.960($8|0$)&1.000($0|0$)&0.990($1|1$)\\
			\hline
		\end{tabular*}
	\end{table*}
	
	\begin{table*}
		\caption{Relative frequency estimates for $P(\hat{r} = 3)$ and $P(\hat{c} = 3)$
			with 200 reduplicate samples for the AR coefficients with case~(3) and $\bm\Sigma_1$ and $\bm\Sigma_2$ with case~(2) }
		\label{tab6}
		\scriptsize
		\begin{tabular*}{\textwidth}{@{\extracolsep{\fill}}ccccccccccc}
			\hline
			$(\delta,\omega)$& $(0.5,0.5)$ \\
			\hline
			$(p,q)$&$n$&$\quad ER_o$&$\quad SR_o$&$\quad MR_o$&$\quad \widetilde{ER}_o$&$\quad \widetilde{SR}_o$&$\quad \widetilde{MR}_o$\\
			\hline
			(20,20)& $200$ &0.820($35|1$)&0.975($5|0$)&0.940($12|0$)&0.785($42|1$)&0.965($7|0$)&0.930($14|0$)\\
			& $800$ &0.845($31|0$)&0.990($12|0$)&0.935($13|0$)&0.770($46|0$)&0.950($10|0$)&0.935($13|0$)\\
			\hline
			(20,40)& $200$ &0.810($38|0$)&0.925($15|0$)&0.925($15|0$)&0.825($35|0$)&0.990($2|0$)&0.905($19|0$)\\
			& $800$ &0.850($30|0$)&0.950($10|0$)&0.925($15|0$)&0.825($35|0$)&0.975($5|0$)&0.930($14|0$)\\
			\hline
			(40,40)& $200$ &0.845($31|0$)&0.970($6|0$)&0.895($21|0$)&0.825($35|0$)&0.975($5|0$)&0.915($16|1$)\\
			& $800$ &0.825($35|0$)&0.985($3|0$)&0.905($19|0$)&0.825($35|0$)&0.995($1|0$)&0.935($13|0$)\\
			\hline
			$(p,q)$&$n$&$\quad ER$&$\quad SR$&$\quad MR$&$\quad \widetilde{ER}$&$\quad \widetilde{SR}$&$\quad \widetilde{MR}$\\
			\hline
			(20,20)& $200$ &0.820($33|3$)&0.990($2|0$)&0.950($8|2$)&0.810($37|1$)&0.985($2|1$)&0.930($14|0$)\\
			& $800$ &0.850($30|0$)&0.995($1|0$)&0.950($10|0$)&0.785($43|0$)&0.975($5|0$)&0.940($12|0$)\\
			\hline
			(20,40)& $200$ &0.820($36|0$)&0.965($7|0$)&0.950($10|0$)&0.835($32|1$)&0.995($1|0$)&0.910($16|2$)\\
			& $800$ &0.855($29|0$)&0.975($5|0$)&0.955($9|0$)&0.830($32|2$)&1.000($0|0$)&0.930($14|0$)\\
			\hline
			(40,40)& $200$ &0.845($31|0$)&0.985($3|0$)&0.910($18|0$)&0.840($32|0$)&0.995($1|0$)&0.945($10|1$)\\
			& $800$ &0.835($33|0$)&1.000($3|0$)&0.920($16|0$)&0.860($26|2$)&0.995($1|0$)&0.940($11|1$)\\
			\hline
			$(\delta,\omega)$& $(0,0.5)$ \\
			\hline
			$(p,q)$&$n$&$\quad ER_o$&$\quad SR_o$&$\quad MR_o$&$\quad \widetilde{ER}_o$&$\quad \widetilde{SR}_o$&$\quad \widetilde{MR}_o$\\
			\hline
			(20,20)& $200$ &0.935($12|1$)&0.995($1|0$)&0.990($2|0$)&0.940($12|0$)&1.000($0|0$)&0.990($2|0$)\\
			& $800$ &0.970($5|1$)&1.000($0|0$)&0.985($3|0$)&0.945($11|0$)&0.990($1|1$)&0.985($2|1$)\\
			\hline
			(20,40)& $200$ &0.940($12|0$)&0.990($2|0$)&0.985($3|0$)&0.900($19|1$)&1.000($0|0$)&0.975($4|1$)\\
			& $800$ &0.960($8|0$)&0.995($1|0$)&0.990($2|0$)&0.950($10|0$)&1.000($0|0$)&0.980($3|1$)\\
			\hline
			(40,40)& $200$ &0.940($11|1$)&0.995($1|0$)&0.985($2|1$)&0.950($10|0$)&0.990($2|0$)&0.985($3|0$)\\
			& $800$ &0.955($9|0$)&0.995($0|0$)&0.990($2|0$)&0.955($9|0$)&0.995($1|0$)&0.990($2|0$)\\
			\hline
			$(p,q)$&$n$&$\quad ER$&$\quad SR$&$\quad MR$&$\quad \widetilde{ER}$&$\quad \widetilde{SR}$&$\quad \widetilde{MR}$\\
			\hline
			(20,20)& $200$ &0.930($12|2$)&0.995($1|0$)&0.990($2|0$)&0.940($9|3$)&1.000($0|0$)&0.990($2|0$)\\
			& $800$ &0.975($4|1$)&1.000($0|0$)&0.985($3|0$)&0.940($10|2$)&0.995($0|1$)&0.985($2|1$)\\
			\hline
			(20,40)& $200$ &0.945($11|0$)&0.995($1|0$)&0.990($1|0$)&0.910($16|2$)&1.000($0|0$)&0.970($3|3$)\\
			& $800$ &0.965($7|0$)&1.000($0|0$)&0.990($1|1$)&0.950($10|0$)&1.000($0|0$)&0.980($2|2$)\\
			\hline
			(40,40)& $200$ &0.940($11|1$)&1.000($0|0$)&0.985($2|1$)&0.950($10|0$)&1.000($0|0$)&0.980($2|2$)\\
			& $800$ &0.975($5|0$)&1.000($0|0$)&0.990($2|0$)&0.955($9|0$)&0.995($1|0$)&0.990($2|0$)\\
			\hline
			$(\delta,\omega)$& $(0,0)$ \\
			\hline
			$(p,q)$&$n$&$\quad ER_o$&$\quad SR_o$&$\quad MR_o$&$\quad \widetilde{ER}_o$&$\quad \widetilde{SR}_o$&$\quad \widetilde{MR}_o$\\
			\hline
			(20,20)& $200$ &0.985($2|1$)&0.995($1|0$)&0.990($2|0$)&0.980($4|0$)&1.000($0|0$)&1.000($0|0$)\\
			& $800$ &0.970($5|1$)&0.995($1|0$)&0.995($1|0$)&0.970($5|1$)&0.995($1|0$)&0.995($1|0$)\\
			\hline
			(20,40)& $200$  &0.985($2|1$)&1.000($0|0$)&1.000($0|0$)&0.985($2|1$)&1.000($0|0$)&0.990($1|1$)\\
			& $800$ &0.990($2|0$)&1.000($1|0$)&0.995($1|0$)&0.975($1|2$)&1.000($0|0$)&1.000($0|0$)\\
			\hline
			(40,40)& $200$ &0.995($1|0$)&1.000($0|0$)&1.000($0|0$)&0.990($2|0$)&1.000($0|0$)&0.995($0|1$)\\
			& $800$ &0.990($2|0$)&1.000($0|0$)&1.000($0|0$)&0.995($0|0$)&1.000($0|0$)&1.000($0|0$)\\
			\hline
			$(p,q)$&$n$&$\quad ER$&$\quad SR$&$\quad MR$&$\quad \widetilde{ER}$&$\quad \widetilde{SR}$&$\quad \widetilde{MR}$\\
			\hline
			(20,20)& $200$ &0.985($1|2$)&0.995($1|0$)&0.990($2|0$)&0.980($4|0$)&1.000($0|0$)&1.000($0|0$)\\
			& $800$ &0.975($1|2$)&1.000($0|0$)&1.000($0|0$)&0.970($4|2$)&0.995($1|0$)&0.995($1|0$)\\
			\hline
			(20,40)& $200$ &0.985($2|1$)&1.000($0|0$)&1.000($0|0$)&0.985($2|3$)&1.000($0|0$)&0.995($0|1$)\\
			& $800$ &0.990($1|1$)&1.000($0|0$)&0.995($1|0$)&0.970($1|5$)&1.000($0|0$)&1.000($0|0$)\\
			\hline
			(40,40)& $200$ &0.995($1|0$)&1.000($0|0$)&1.000($0|0$)&0.990($2|0$)&1.000($0|0$)&1.000($0|0$)\\
			& $800$ &0.995($1|0$)&1.000($0|0$)&1.000($0|0$)&0.995($1|0$)&1.000($0|0$)&1.000($0|0$)\\
			\hline
		\end{tabular*}
	\end{table*}

	\newpage

\end{document}